\documentclass{aa}  

\usepackage{graphicx}
\usepackage{txfonts}
\usepackage{mhchem}
\usepackage{bm}
\usepackage{comment}
\usepackage[small,bf]{caption}
\usepackage[subrefformat=parens]{subcaption}
\def\au{{\rm\, au}}
\def\msol{{\rm M_\odot}}

\def\rd0{r_{\rm d,0}}
\def\ga{\ \lower 3pt\hbox{${\buildrel > \over \sim}$}\ }
\def\la{\ \lower 3pt\hbox{${\buildrel < \over \sim}$}\ }

\DeclareGraphicsExtensions{.pdf,.png,.jpg}

\begin{document} 
   \title{Water delivery by pebble accretion to rocky planets in habitable zones in evolving disks}
   \author{Shigeru Ida \inst{1},
          Takeru Yamamura \inst{2}           
          \and Satoshi Okuzumi \inst{2}}
   \institute{Earth-Life Science Institute, Tokyo Institute of Technology, Meguro-ku, Tokyo 152-8550, Japan  \\  \email{ida@elsi.jp}
             \and
   Department of Earth and Planetary Sciences, Tokyo Institute of Technology, Meguro-ku, Tokyo 152-8551, Japan \\
  }
   \date{DRAFT:  \today}


\abstract{
The Earth's ocean mass is only $2.3 \times 10^{-4}$ of the whole planet mass. 
Even including water in the interior, the water fraction would be at most $10^{-3}-10^{-2}$.
Ancient Mars may have had a similar or slightly smaller water fraction.
It has not been clear what controlled the amount of water in these planets,
although several models have been proposed.
It is important to clarify the control mechanism to discuss water delivery
to rocky planets in habitable zones in exoplanetary systems,
as well as that to the Earth and Mars in our Solar system.
}
{
Here, we consider water delivery to planets by icy pebbles after the snowline inwardly passes
the planetary orbits.
We derive the water mass fraction ($f_{\rm water}$) of the final planet as a function of disk parameters
and discuss the parameters that reproduce a small value of $f_{\rm water}$ comparable to that inferred for the Earth and ancient Mars.
}
{
We calculate the growth of icy dust grains to pebbles and the pebble 
radial drift with a 1D model, by simultaneously solving the snowline migration and
dissipation of a gas disk.
With the obtained pebble mass flux, we calculate accretion of icy pebbles
onto planets after the snowline passage to evaluate $f_{\rm water}$ of the planets.
}
{
We find that $f_{\rm water}$ is regulated by the total mass ($M_{\rm res}$) of 
icy dust materials preserved in the outer disk regions at the timing
($t = t_{\rm snow}$) of the snowline passage of the planetary orbit.
Because $M_{\rm res}$ decays rapidly
after the pebble formation front reaches the disk outer edge (at $t = t_{\rm pff}$),
$f_{\rm water}$ is sensitive to 
the ratio $t_{\rm snow}/t_{\rm pff}$, which is determined by the disk parameters.
We find $t_{\rm snow}/t_{\rm pff} < 10$ or $> 10$ is important.
Analytically evaluating $M_{\rm res}$,
we derive an analytical formula of $f_{\rm water}$ that reproduces the numerical results.
}
{
Using the analytical formula, we find that
$f_{\rm water}$ of a rocky planet near 1 au is
similar to the Earth, $\sim 10^{-4}-10^{-2}$,
in the disks with initial disk size of 30-50 au and
the initial disk mass accretion rate of $\sim (10^{-8}-10^{-7}) M_\odot{\rm/r}$
for disk depletion timescale of $\sim$ a few M years. 
Because these disks may be median or slightly compact/massive disks,
our results suggest that the water fraction of rocky planets in habitable zones 
may be often similar to that of the Earth, if the icy pebble accretion is responsible for
the water delivery.
}
\titlerunning{Water delivery by pebble accretion}
\maketitle
%

\section{Introduction}
Earth-size planets are being discovered in habitable zones in exoplanetary systems.
Habitable zones (HZs) are defined as a range of orbital radius,
in which liquid water can exist on the planetary surface,
{\it if H$_2$O exist there.}
However, as long as equilibrium temperature is concerned, 
H$_2$O ice grains condense only well beyond the habitable zones,
because the gas pressure of protoplanetary disks is many orders of
magnitude lower than that of planetary atmosphere and the condensation
temperature is considerably lower than that at 1 atm.
Hereafter, we simply call H$_2$O in solid/liquid phase as ``water."
For the Earth, volatile supply by the gas capture from the disk is ruled out, 
because observed values of rare-Earth elements are too low
in the Earth to be consistent with the disk gas capture \citep[e.g.,][]{Brown49}. 
Therefore, the water in the Earth would have been
delivered from the outer regions of the disk during planet formation.

One possible water delivery mechanism to the Earth
is inward scattering of water-bearing asteroids by Jupiter (e.g., \citealt{Raymond2004}).
If this is a dominant mechanism of water delivery,
the amount of delivered water is rather stochastic and depends on 
configurations of giant planets in the planetary systems. 
If water is not delivered, a rocky planet in a HZ may not be able to be an actual habitat. 
On the other hand, too much water makes a planet without continents,
where nutrients supply may not be as effective as in the Earth.
The ocean of the Earth comprises only 0.023 wt.\% and
such a right amount enables ocean and continents to coexist.
The mantle may preserve water in the transition zone with a comparable amount of the ocean
(e.g., \citealt{BandK2003}; \citealt{Hirschmann2006}; \citealt{Fei2017}), 
while the core could have H equivalent to 2 wt.\% of H$_2$O of the Earth \citep{Nomura2014}.
However, the original water fraction of the Earth would still be very small ($\sim 10^{-4}-10^{-2}$), 
even with the possible water reservoir in the interior, 
because neither stellar irradiation at $\sim 1$ au \citep{MachidaAbe2010} nor
giant impacts \citep{GendaAbe2005} can vaporize the majority of the water from the Earth's gravitational potential. 
Note that it is inferred that Mars may have subsurface water of 
$10^{-4} -10^{-3}$ of Mars mass 
(e.g., \citealt{AchilleHynek2010}; \citealt{Clifford2010}; \citealt{Kurokawa2014}).
From the high D/H ratio observed in the Venus atmosphere, the early Venus may also have had ocean of the fraction of $10^{-5}-10^{-3}$ and lost the H$_2$O vapor through runaway greenhouse effect \citep[e.g.,][]{Donahue1982,Greenwood2018}.  
The order of water fraction looks similar at least between the Earth and the ancient Mars. 
Although the estimated total mass fraction of water in the Earth and Mars 
has relatively large uncertainty ranging from $10^{-4}$ to $10^{-2}$, 
the range is still much smaller than the dispersion predicted by
the water-bearing asteroid collision model,
which ranges from $10^{-5}$ to $10^{-1}$, depending on
the formation timing, the history of orbital migration/eccentricity evolution of
gas giant planets and  the original surface density of planetesimals 
\citep[e.g.,][]{Morbidelli2000, Lunine2003, Raymond2004, OBrien2014, Matsumura2016}.
It is not clear if the similar orders of water fraction between the Earth and Mars 
is just a coincidence.

\cite{Sato2016} investigated water delivery by icy pebble accretion.
The pebble accretion has been proposed as a new mode of
planet accretion \citep[e.g.,][]{OrmelKlahr2010,OrmelKlahr2012}. 
Radiative transfer calculations for viscous accretion disk models show 
that the water snowline at $\sim 170$ K may migrate to inside of 1 au
with the grain opacity of $\ga {\rm mm}$ size, 
when the disk accretion rate is $\dot{M}_{\rm g} \la 10^{-8} M_\odot/{\rm y}$
(e.g., \citealt{GaraudLin2007}; \citealt{Min2011}; \citealt{Oka2011}),
which is a typical value of $\dot{M}_{\rm g}$ of classical T-Tauri stars \citep{Hartmann1998}.
After the snowline inwardly passes a planetary orbit,
icy pebbles can be accreted by the planet.
In situ ice condensation near the planet orbit is unlikely,
because the disk gas there was once in outer region before it migrates to the inner region
and icy components have been already condensed to icy grains 
and subtracted in the outer region \citep{Morbidelli2016, Sato2016}.
\cite{Sato2016} calculated the time evolution of icy pebble mass flux,
the solid surface density in the disk, and the growth of a hypothetical planet
at 1 au by icy pebble accretion with a 1D model.
They found that the pebble accretion is so efficient
that the water fraction of the planet rapidly increases
after the snowline passage.
They assumed a static disk and 
artificially set the timings of snowline passage and removal of the disk
that truncates pebble accretion.
They found that the water mass fraction of the final planet
is very sensitive to these timings
and it is zero or more than 0.1 in many cases.
The modest water mass fraction of $10^{-4}-10^{-2}$
is possible only if the disk is compact ($< 100 \,{\rm au}$)
and the snowline passage at 1 au later than 2--4 My after icy dust growth starts,
which could be a narrow window of the parameters.

The sensitive dependence of the final water fraction
requires that the snowline migration and the decay of the icy dust surface density
must be consistently calculated in an evolving disk.
Here we use the disk evolution model based on
the self-similar solution for accretion disks
with constant viscosity parameter $\alpha$ \citep{LyndenPringle1974}\footnote{The wind-driven disk accretion that is recently proposed \citep[e.g.,][]{Suzuki16,Bai16}
is commented on in section 2.1}.
The snowline migration, the disk gas decay, and
growth/drift of pebbles and the associated evolution of the icy dust surface density
are simultaneously calculated by a 1D disk evolution model.
The growth and drift of pebbles are tracked using the 
single-size approximation formulated by \citet{Ormel2014} and \citet{Sato2016},
which enables us to perform fast calculations and survey broad ranges of parameters. 
Using the numerical results, we will also derive an analytical
formula for the final water mass fraction of the planets
determined by the disk model parameters.

In section 2, we describe the calculation model that we used.
In section 3, the results of numerical simulation are shown.
We derive the semi-analytical formula that successfully reproduces the numerical results in section 4.
In section 5, using the analytical formula, we study
the dependence of the planetary water fraction on the disk and pebble accretion parameters,
and discuss the disk parameters to realize the water fraction of $10^{-4}-10^{-2}$, which 
corresponds to the present Earth and ancient Mars.
We will show that the disk parameters are not in a narrow window and
are rather realized in modest disks.
Sections 6 and 7 are discussion and summary.

\section{Method}

\subsection{Gas disk model}

In general, an accretion disk consists of an inner region where the viscous heating is dominated
and an outer region where irradiation from the host star is dominated.
According to \citet{Ida-et2016},
we set the disk mid-plane temperature for the 
viscous-heating dominated region ($T_{\rm vis}$) and the irradiation dominated region ($T_{\rm irr}$) to be 
\begin{align}
\label{eq:tempvis}
&T_{\rm vis} \simeq 130 \left(\frac{\alpha}{10^{-2}} \right)^{-1/5} 
\left(\frac{M_{\ast}}{M_{\odot}} \right)^{3/10}
\left(\frac{\dot{M}_{\rm g}}{10^{-8} M_{\odot} / {\rm y}} \right)^{2/5} \left(\frac{r}{1 {\rm au}} \right)^{-9/10} \, {\rm K}, \\
&T_{\rm irr} \simeq 130 \left(\frac{L_{\ast}}{L_{\odot}} \right)^{2/7} \left(\frac{M_{\ast}}{M_{\odot}} \right)^{-1/7} \left(\frac{r}{1 {\rm au}} \right)^{-3/7}\, {\rm K},
\label{eq:tempirr}
\end{align}
where $r$ is the distance from the host star, 
$\dot{M}_{\rm g}$ is the disk gas accretion rate, which is almost independent of $r$
except in outermost region, 
$L_{\ast}$ and $M_{\ast}$ are respectively the luminosity and mass of the host star, 
and $L_{\odot}$ and $M_{\odot}$ are their values of the Sun.
We adopt the alpha prescription for the disk gas turbulent viscosity \citep{Shakura1973},
$\nu \simeq \alpha h_{\rm g}^2 \Omega$, where $h_{\rm g}$ is the disk gas scale height,
defined by $h_{\rm g} = c_s/\Omega$, $c_s$ and $\Omega$ are respectively the sound velocity and
Kepler frequency, and $\alpha$ $(<1)$ is a parameter to represent the strength of the turbulence.
We here use slightly lower $T_{\rm irr}$ than that in \cite{Ida-et2016},
assuming lower opacity 
with mm-sized dust grains \citep{Oka2011}, because we consider relatively inner disk regions near the snowline and pebbles there are those which have grown in outer regions and drifted inward.
If micron-sized grains are assumed, the same temperature is
realized with about ten times smaller $\dot{M}_{\rm g}$.

One fundamental assumption behind Equation~(\ref{eq:tempvis}) is that 
the rate of viscous heating per unit volume scales linearly with the gas density, 
and is therefore highest at the midplane. 
This assumption is questioned by magnetohydrodynamic (MHD) models of protoplanetary disks, 
which show that accretion heating dominantly takes place on the disk surface 
\citep{HiroseTurner11}.  
Recently, Mori et al. (2019, submitted) investigated this issue using a series of MHD simulations 
including all non-ideal MHD effects, finding that the midplane temperature derived 
from the simulations is generally lower than 
that from Equation~(\ref{eq:tempvis}) because the heat generated near the disk surface can 
easily be lost through radiation.  
Mori et al. (2019) also find that MHD disk winds, which are not included in our disk model, 
take away $\approx 30\%$ of the magnetic energy that would be available for disk heating 
if the winds were absent.   
Although there are disk evolution models accounting for the mass and angular momentum
loss due to MHD disk winds \citep[e.g.,][]{Armitage2013,Suzuki16,Bai16,Hasegawa2017}, 
none of them take into account the two effects mentioned above. 
For this reason, we opt to adopt the more classical viscous disk model in this study. 
We note that the viscous accretion model will serve as a good approximation 
of real protoplanetary disks if some hydrodynamical instabilities drive turbulence near the midplane
\citep[see][for recent reviews on hydrodynamic instabilities of protoplanetary disks]{LyraUmurhan18,Klahr18}.

The gas disk aspect ratios corresponding to Eqs.~(\ref{eq:tempvis}) and (\ref{eq:tempirr}) are
\begin{align}
\label{eq:hvis}
&\frac{h_{\rm g,vis}}{r} \simeq 0.022 \left(\frac{\alpha}{10^{-2}} \right)^{-1/10} 
\left(\frac{M_{\ast}}{M_{\odot}} \right)^{-7/20}
\left(\frac{\dot{M}_{\rm g}}{10^{-8} M_{\odot} / {\rm y}} \right)^{1/5} \left(\frac{r}{1 {\rm au}} \right)^{1/20}, \\
\label{eq:hirr}
&\frac{h_{\rm g,irr}}{r} \simeq 0.022 \left(\frac{L_{\ast}}{L_{\odot}} \right)^{1/7} \left(\frac{M_{\ast}}{M_{\odot}} \right)^{-4/7} \left(\frac{r}{1 {\rm au}} \right)^{2/7},
\end{align}
where $h_{\rm g,vis}$ and $h_{\rm g,irr}$ are the gas scale hight in the
viscous-heating dominated and irradiation dominated regions, respectively.
Hereafter, we perform simulations with $L_{\ast}=L_\odot$ and $M_{\ast}=M_\odot$,
while we retain their dependences in the formulas. 
The disk region is viscous-heating dominated if $T_{\rm vis} > T_{\rm irr}$.
Otherwise, the irradiation dominates.
The transition radius between the viscous-heating and irradiation dominated regions is given by
\begin{equation}
\label{eq:rvisirr}
r_{\rm vis/irr} \simeq 1 \left(\frac{\alpha}{10^{-2}} \right)^{-14/33}  \left(\frac{L_{\ast}}{L_{\odot}} \right)^{-20/33} \left(\frac{M_{\ast}}{M_{\odot}} \right)^{31/33} \left(\frac{\dot{M}_{\rm g}}{10^{-8} M_{\odot} / {\rm y}} \right)^{28/33} \, {\rm au}.
\end{equation}

Defining the snowline by the location at $\sim 170$ K,
$r_{\rm snow} \simeq \max(r_{\rm snow,vis}, r_{\rm snow,irr})$, where
\begin{equation}
r_{\rm snow,vis} \simeq 0.74 \left(\frac{M_*}{M_\odot}\right)^{1/3}
\left(\frac{\alpha}{10^{-2}}\right)^{-2/9}
\left(\frac{\dot{M}_{\rm g}}{10^{-8} M_{\odot} / {\rm y} }\right)^{4/9} {\rm au},
\label{eq:r_snow_vis}
\end{equation}
\begin{equation}
r_{\rm snow,irr} \simeq 0.53 \left(\frac{L_{*}}{L_\odot}\right)^{2/3}\left(\frac{M_{*}}{M_\odot}\right)^{-1/3} {\rm au}.
\label{eq:r_snow_irr}
\end{equation}
As $\dot{M}_{\rm g}$ decreases with time, the snowline
migrates inward in the viscous-heating dominated region
until $r_{\rm snow,vis}$ becomes equal to $r_{\rm snow,irr}$
($\dot{M}_{\rm g} \ga 5\times 10^{-9} M_{\odot} / {\rm y}$ for $\alpha=10^{-2}$).
For calculating the evolution of the pebble flux, it may be enough to
set up a static disk distribution as 
\cite{Sato2016} did.   
However, in order to describe the snowline migration and disk gas depletion
(which determines timings of start and termination of the icy pebble supply), 
we need an evolving gas disk model.

Specific orbital angular momentum is proportional to
a square root of orbital radius $r$ and
most of disk mass exists in the outer irradiation dominated region.
Angular momentum transfer
in the entire disk that determines the snowline migration and 
the entire disk gas depletion is regulated by the evolution of the outer disk region.
The region near the snowline is not a hot region where
the viscous heating is significantly higher than the irradiation heating (Eqs.~\ref{eq:rvisirr} and \ref{eq:r_snow_vis}).
So, for our purpose, the entire disk evolution model can be approximated by
a irradiation dominated disk.
In the irradiation dominated disk,
the viscosity $\nu \propto \alpha T r^{3/2} \propto \alpha r ^{15/14}$ (Eq.~\ref{eq:tempirr}).
Because it is similar to the disk with the viscosity $\nu \propto r$ with a constant $\alpha$,
we adopt the self-similar solution with $\nu \propto r$
by \cite{LyndenPringle1974} for the dynamical evolution of the entire disk,
while we take into account the snowline evolution regulated by time evolution of the viscous heating (Eq.~\ref{eq:r_snow_vis}).  

In the self-similar solution, well inside the initial characteristic disk size ($r_{\rm d,0}$),
beyond which the surface density decays exponentially, the disk accretion rate is given 
as a function of time by
\begin{align}
\dot{M}_{\rm g} & \simeq 3\pi \Sigma_{\rm g} \nu 
\simeq \dot{M}_{\rm g,0} \tilde{t}^{\; -3/2},
\label{eq:flux1}
\end{align}
where $\Sigma_{\rm g}$ is the disk gas surface density, 
$\nu$ is the effective viscosity at $r$, 
$\tilde{t} = 1 + t/t_{\rm diff}$, 
$t_{\rm diff} = r_{\rm d,0}^2/3\nu_0 = r_{\rm d,0} r /3\nu$ (where $\nu_0$ is the viscosity at $r_{\rm d,0}$),
and $\dot{M}_{\rm g,0}$ is the initial disk accretion rate, respectively. 
Inversely, the time evolution of the surface gas density $\Sigma_{\rm g}$ is given by
\begin{align}
 \Sigma_{\rm g} & \simeq \frac{\dot{M}_{\rm g,0}}{3\pi \nu} \tilde{t}^{\; -3/2}
 \exp{\left( - \frac{r}{\tilde{t}\,r_{\rm d,0}} \right)}, \label{eq:Sigma_g2} 
\end{align}
where we included the time-dependent exponential taper
in the full form of the self-similar solution,
because we will need to evaluate the total disk mass. 
Because $\nu \propto r$, $\Sigma_{\rm g}\propto 1/r$ for $r \ll \rd0$.

We add the effect of the photoevaporation with the rate $\dot{M}_{\rm pe}$,
although the standard self-similar solution does not have such a term (also see section 3.1).
We are concerned with the region near the snowline.
We assume that $r \ll r_{\rm d,0}$ and
the photoevaporation occurs mainly in the outer region
with the constant rate of $\dot{M}_{\rm pe}$.
{\bf Accordingly, we set 
\begin{align}
\dot{M}_{\rm g} & = \dot{M}_{\rm g,0} \tilde{t}^{\; -3/2}
\exp{\left( - \frac{r}{\tilde{t}\, r_{\rm d,0}} \right)} - \dot{M}_{\rm pe}, \label{eq:flux20} \\
\Sigma_{\rm g} & \simeq \frac{\dot{M}_{\rm g,0} \tilde{t}^{\; -3/2}
\exp{\left( - \frac{r}{\tilde{t}\, r_{\rm d,0}} \right)}
-\dot{M}_{\rm pe}}{3\pi \nu}.
\label{eq:Sigma_g30}
 \end{align}
 While we use Eqs.~(\ref{eq:flux20}) and (\ref{eq:Sigma_g30}) for the numerical simulation,
 we use approximate forms, which are easier to be analytically treated,
 for derivation of analytical formulas, as follows:  
 \begin{align}
\dot{M}_{\rm g} & = \dot{M}_{\rm g,0} \tilde{t}^{\; -3/2} - \dot{M}_{\rm pe}, \label{eq:flux2} \\
\Sigma_{\rm g} & \simeq \frac{\dot{M}_{\rm g,0} \tilde{t}^{\; -3/2}-\dot{M}_{\rm pe}}{3\pi \nu} 
 \exp{\left( - \frac{r}{\tilde{t}\, r_{\rm d,0}} \right)}.  
\label{eq:Sigma_g3}
\end{align}
Because $\dot{M}_{\rm g}$ and $\Sigma_{\rm g}$ are
overestimated with Eqs.~(\ref{eq:flux2}) and (\ref{eq:Sigma_g3}) 
only in the exponential tail regions, the approximation 
does not introduce significant errors in the analytical formulas.
\footnote{\bf In the published version in A\&A, we started
from Eqs.~(\ref{eq:flux2}) and (\ref{eq:Sigma_g3}) without
explaining the approximation from Eqs.~(\ref{eq:flux20}) and (\ref{eq:Sigma_g30}).}}
Integrating Eq.~(\ref{eq:Sigma_g3}), the disk mass at $t$ is given by 
\begin{equation}
 M_{\rm g}(\tilde{t}) = \int 2\pi r \Sigma_{\rm g}(\tilde{t}) \, dr = 2 \, \tilde{t} \, t_{\rm diff}(\dot{M}_{\rm g,0} \tilde{t}^{\; -3/2} - \dot{M}_{\rm pe}), \label{eq:M_g2} 
\end{equation}
where we used $r/3 \nu=t_{\rm diff}/\rd0$.
The above equations show that
the disk accretion rate $\dot{M}_{\rm g}$ and the disk gas surface density $\Sigma_{\rm g}$
quickly decay 
when $\dot{M}_{\rm g}$ decreases to the level of $\dot{M}_{\rm pe}$.
This photoevaporation effect avoids long-tail existence of disk gas significantly longer than a few My.

Here, we describe the disks with the parameters, $t_{\rm diff}$,
$\dot{M}_{\rm g,0}$, $\dot{M}_{\rm pe}$ and $r_{\rm d,0}$.
The disk depletion timescale ($t_{\rm dep}$) 
and the disk gas accretion rate onto the host star ($\dot{M}_{\rm g}$)
are better constrained by observations than the other parameters:
$t_{\rm dep} \sim 10^6-10^7$ y and
$\dot{M}_{\rm g} \sim (10^{-9}-10^{-7}) M_\odot/{\rm y}$
for solar-type stars
(e.g., \citealt{Haisch2001}; \citealt{Hartmann1998}; \citealt{WilliamsCieza2011}; \citealt{Hartmann2016}).
We focus ourselves on the systems around solar-type stars.
We assume that angular momentum transfer by turbulent viscous diffusion
is a major mechanism for
the disk depletion, rather than photoevaporation.
We identify $t_{\rm dep}$ as $t_{\rm diff}$.
We use $\dot{M}_{\rm g,0}$ may be slightly higher than the observed 
averaged values of $\dot{M}_{\rm g}$, because we want to set the snowline beyond the 
orbits of planetary embryos.
We perform simulations with
$t_{\rm diff} = 10^6$ y and $3 \times 10^6$ y, and
$\dot{M}_{\rm g,0} = 3\times 10^{-8} M_\odot/{\rm y}$ and $10^{-7} M_\odot/{\rm y}$.
Although $r_{\rm d,0}$ is not observationally well constrained,
\cite{Sato2016} showed that this parameter is the most important for  
the time evolution of the pebble mass flux.
For $r_{\rm d,0}$, we use $r_{\rm d,0}=30\au, 100\au ,$ and 300 au,
which correspond to the range of the observationally inferred disk size \citep{WilliamsCieza2011}. 
We adopt the photoevaporation rate $\dot{M}_{\rm pe}= 10^{-9}M_\odot/{\rm y}$ and $10^{-8} M_\odot/{\rm y}$.

The disk diffusion timescale is
\begin{align}
t_{\rm diff} & \simeq \frac{r^2_{\rm d.0}}{3 \nu_0} 
= \frac{r_1 r_{\rm d.0}}{3 \nu_1}
= \frac{r_1^2}{3 \alpha h^2_{\rm g,1} \Omega_{\rm K,1}} \frac{r_{\rm d.0}}{r_1} 
\nonumber\\
 & = \frac{1}{6 \pi \alpha} \left( \frac{r_1}{h_{\rm g,1}} \right)^2 \frac{r_{\rm d.0}}{r_1}\; {\rm y}
\simeq 10^2 \alpha^{-1} \left( \frac{r_{\rm d.0}}{1\,{\rm au}} \right) \; {\rm y},
\label{eq:alpha_derive}
\end{align}
where subscript "1" expresses values at 1au, 
$\nu_{\rm d.0} = \nu(r_{\rm d.0})$ is viscous coefficient at the outer edge of the disk,
and we used $\nu = \nu_1 \cdot (r / r_1)$. 
From this equation, $\alpha$ is calculated as
\begin{align}
\alpha \simeq 10^{-2} \left( \frac{t_{\rm diff}}{10^6\; {\rm y}} \right)^{-1} \left( \frac{r_{\rm d.0}}{100 \au} \right).
\label{eq:alpha}
\end{align}
Note that the value of $\alpha$ that we use in our simulations depends on the choice of 
the parameters $t_{\rm diff}$ and $r_{\rm d,0}$, but it is within a reasonable range, $\alpha = 10^{-3}-3\times 10^{-2}$.
In our formulation for disks, we prefer the setting of
the observable valuables, $t_{\rm diff}$ and $\rd0$,
within the observationally inferred ranges to 
a simple assumption of a constant $\alpha$.

\begin{figure*}\begin{center} 
\includegraphics[width=150mm]{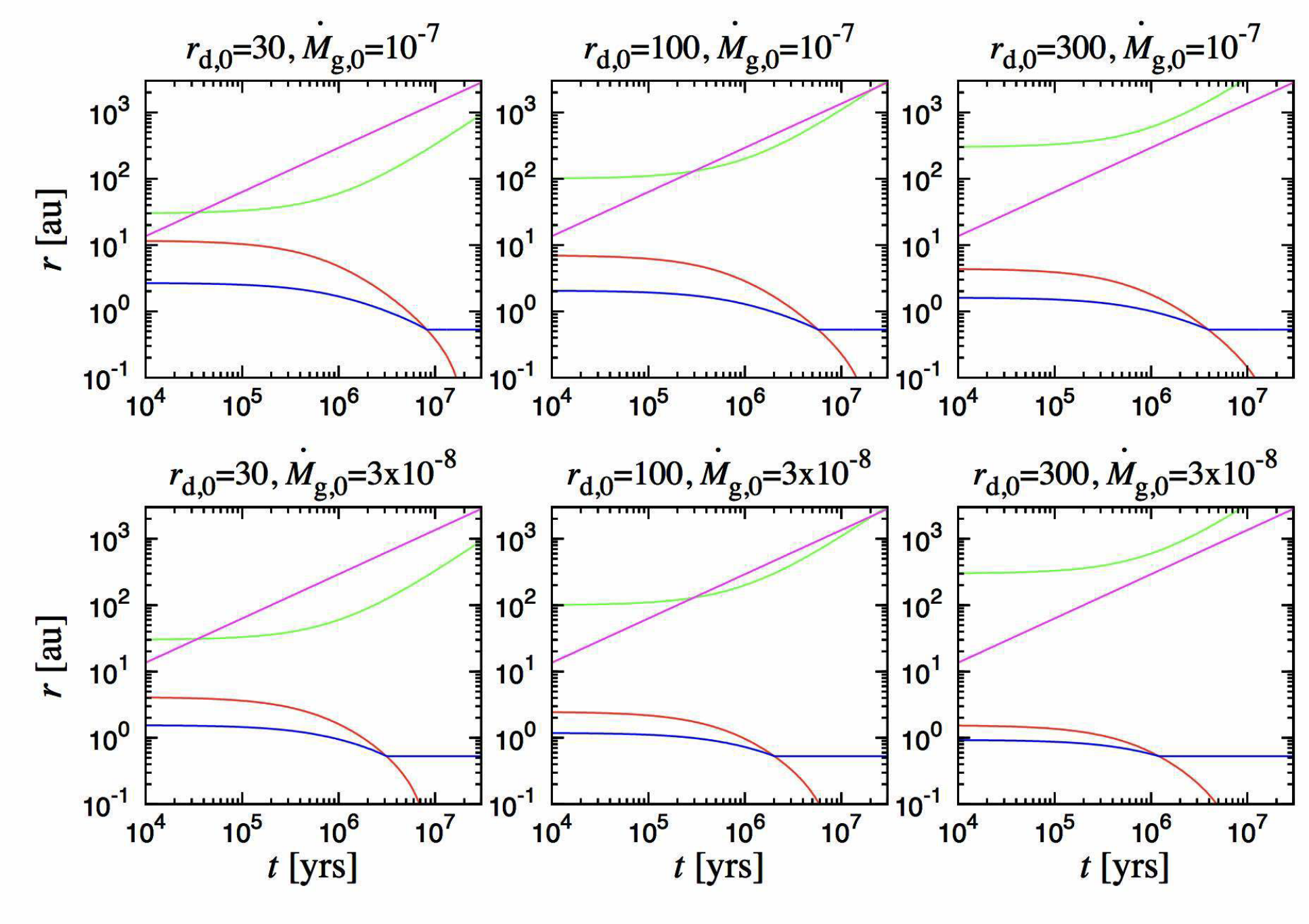}
\end{center}
\caption{
The time evolution of 
the transition radius $r_{\rm vis/irr}$ (Eq.~\ref{eq:rvisirr})
with the red lines, 
the snowline $r_{\rm snow} = \max(r_{\rm snow,vis}, r_{\rm snow,irr})$
\; (Eqs.~\ref{eq:r_snow_vis} and \ref{eq:r_snow_irr})
with the blue lines,
the disk characteristic radius $\tilde{t} \rd0$
with the green lines, and
the pebble formation front $r_{\rm pff}$ (Eq.~\ref{eq:tpff}) with the magenta lines,
for various disk evolution parameters.
The parameters, $\rd0$ in unit of au 
and $\dot{M}_{\rm g,0}$ in unit of $M_\odot/{\rm y}$, are labeled on the individual panels.
The other parameters are the same for the four panels:
$\dot{M}_{\rm pe}=10^{-9}M_\odot/{\rm y}$
and $t_{\rm diff}=10^6{\rm y}$.
}\label{fig:lines}
\end{figure*}

In Figure~\ref{fig:lines}, we show time evolution of 
the transition radius $r_{\rm vis/irr}$, the snowline $r_{\rm snow}$,
the disk characteristic radius $\tilde{t} \rd0$, and
the pebble formation front $r_{\rm pff}$,
which will be derived in section 3.1 as Eq.~(\ref{eq:tpff}) ,
for six typical parameters of disk evolution.
In general, $r_{\rm pff}$ in magenta lines evolves faster than 
$\tilde{t} \rd0$ in green lines and the pebble flux quickly decays after
 $r_{\rm pff}$ exceeds $\tilde{t} \rd0$, as we will discuss in section 3.1.
 The snowline $r_{\rm snow}$ in blue lines shrinks with time as long as it is 
 in the viscous-heating region (inside of $r_{\rm vis/irr}$ in red lines).

\subsection{Dust/pebble evolution model}

For calculating the pebble mass flux, we adopt the method with the single size approximation 
formulated by 
\cite{Ormel2014} and \cite{Sato2016}. 
The size distribution of icy particles obtained by a full-size simulation is generally peaky at
some size and the peak size depends on $r$. 
In the single size approximation, icy dust particles have a single size that depends on $r$,
corresponding to the peak size.

Here we briefly summarize the dust evolution model (for more details, see \citealt{Sato2016}).
We set the initial particle surface density as $\Sigma_{\rm p} = Z_0 \Sigma_{\rm g}$.
We adopt $Z_0 = 0.01$.
The evolution of the dust/pebble surface density ($\Sigma_{\rm p}$) and 
the peaked mass of the particles ($m_{\rm p}$) are calculated by the equation,
\begin{align}
\label{eq:dust_menpd}
 & \frac{\partial \Sigma_{\rm p}}{\partial t} + \frac{1}{r} \frac{\partial}{\partial r} (r {\rm v}_{r,\rm d} \Sigma_{\rm p}) = 0,\\
\label{eq:dust_cnd}
 & \frac{\partial m_{\rm p}}{\partial t} + {\rm v}_{r,\rm d}  \frac{\partial m_{\rm p}}{\partial r} = \frac{2 \sqrt{\pi} R_{\rm p}^2 \Delta {\rm v}_{\rm pp}}{h_{\rm p}} \Sigma_{\rm p},
\end{align}
where $R_{\rm p} = (3m_{\rm p} / 4 \pi \rho_{\rm int})^{1/3}$ is the particle radius, $\rho_{\rm int}$ is the internal density of icy particles, ${\rm v}_{r, \rm d}$ and $\Delta {\rm v}_{\rm pp}$ are the radial and relative velocities of the particles at the midplane, respectively, and $h_{\rm p}$ is the scale height of the particles given by \citep[e.g.][]{Youdin2007}
\begin{align}
h_{\rm p} = h_{\rm g} \left(1 + \frac{\rm St}{\alpha} \frac{1 + 2{\rm St}}{1 + {\rm St}} \right)^{-1/2}.
\label{eq:pebble_h}
\end{align}
The Stokes number define by ${\rm St} = t_{\rm s} \Omega$ is an important dimensionless number in pebble growth and radial drift,
where $t_{\rm s}$ is the stopping time which represents the timescale of particles's momentum relaxation due to gas drag, given  by
\begin{equation}
t_{\rm s} = \left\{
\begin{array}{lll}
\displaystyle 
\frac{\rho_{\rm int} R_{\rm p}}{\rho_{\rm g} {\rm v}_{\rm th}} & (R_{\rm p} < \frac{9}{4} \lambda_{\rm mfp}; & {\rm Epstein \; regime})  \\
\displaystyle 
\frac{4 \rho_{\rm int} R_{\rm p}^2}{9 \rho_{\rm g} {\rm v}_{\rm th} \lambda_{\rm mfp}} & (R_{\rm p} > \frac{9}{4} \lambda_{\rm mfp}; & {\rm Stokes \; regime})
\end{array}
\right.
\end{equation}
where ${\rm v}_{\rm th} = \sqrt{8 k_{\rm B} T / \pi m_{\rm g}}$ is
the thermal velocity, $\lambda_{\rm mfp}$ is the mean free path of gas particles, 
and $m_{\rm g}$ is gas molecule mass. 
The mean free path is expressed by $\lambda_{\rm mfp} = m_{\rm g} / (\sigma_{\rm mol} \rho_{\rm g})$ where $\sigma_{\rm mol} = 2.0 \times 10^{-15} {\rm cm}^2$ is the molecular collision cross section.

Relative velocity of dust particles $\Delta {\rm v}_{\rm pp}$ is given by
\begin{align}
\Delta {\rm v}_{\rm pp} = \sqrt{(\Delta {\rm v}_{\rm B})^2 + (\Delta {\rm v}_{\rm r})^2 + (\Delta {\rm v}_{\rm \phi})^2 + (\Delta {\rm v}_{\rm z})^2 + (\Delta {\rm v}_{\rm t})^2},
\end{align}
where
$\Delta {\rm v}_{\rm B}, \Delta {\rm v}_{\rm r}, \Delta {\rm v}_{\rm \phi}, \Delta {\rm v}_{\rm z}$, and $\Delta {\rm v}_{\rm t}$ are the relative velocities induced by Brownian motion, radial drift, azimuthal drift, vertical settling, and turbulence, respectively (for detailed expressions, see \cite{Sato2016}).
We assume perfect sticking for $\Delta {\rm v}_{\rm pp} < 30 {\rm m/s}$ (otherwise, we set the right hand side of 
Eq.~(\ref{eq:dust_cnd}) to be zero).

The radial drift velocity of dust particles is 
\begin{align}
{\rm v}_{\rm r,d} = - \frac{2 {\rm St}}{1 + {\rm St}^2} \eta {\rm v}_{\rm K},
\label{eq:vrd}
\end{align}
where ${\rm v}_{\rm K} = r \Omega_{\rm K}$ is the Kepler velocity
and $\eta$ is a deviation of gas rotation velocity from Kepler velocity,
which we set 
\begin{equation}
\eta = 1.1 \times 10^{-3} \left(\frac{r}{1\au}\right)^{1/2}.
\label{eq:eta}
\end{equation}
Because the numerical simulation shows ${\rm St} \ga 0.1$ for migrating pebbles (section 3.1), 
we neglected the effect of the disk gas accretion in Eq.~(\ref{eq:vrd}).
The radial drift timescale is
\begin{align}
t_{\rm drift} \equiv \frac{r}{|{\rm v}_{\rm r,d}|} 
\simeq \frac{1}{2 \eta \, {\rm St} \, \Omega}
\simeq 7 \times 10^4 \left(\frac{\rm St}{0.1}\right)^{-1}
\left(\frac{r}{\rm 100 \,au}\right)
\left(\frac{M_\ast}{M_\odot}\right)^{-1/2}\; {\rm y}.
\label{eq:tmig}
\end{align}
The pebble mass flux through the disk is given by
\begin{equation}
\dot{M}_{\rm peb} = 2 \pi r \Sigma_{\rm p}\, | {\rm v}_{r, \rm d} |
\simeq 4 \pi \,{\rm St} \, \eta r^2 \Sigma_{\rm p} \Omega,
\label{eq:mdot_peb}
\end{equation}
where $\Sigma_{\rm p}$ is the pebble surface density in the pebble migrating region
and we assumed ${\rm St}^2 \ll 1$.

\subsection{Pebble accretion onto planets}

We use the same formulas as \cite{Sato2016} for the pebble accretion rate onto planets,
assuming that the planets are already large enough for pebble accretion in the "settling regime"
(e.g., \citealt{OrmelKlahr2010}; \citealt{Guillot2014}).
\cite{OrmelKlahr2010} derived the cross section of pebble accretion as
\begin{equation}
\pi b_{\rm set}^2 \simeq 4\pi \,{\rm St} \,\frac{GM_{\rm pl}}{\Omega \Delta {\rm v}},
\label{eq:pib2}
\end{equation}
where $M_{\rm pl}$ is the planetary embryo mass and $\Delta {\rm v}$
is the relative velocity between the embryo and pebbles.
The 3D pebble accretion rate $\dot{M}_{\rm pl}$ onto the planetary embryo is given by 
\begin{align}
\dot{M}_{\rm pl} & \simeq \pi b_{\rm set}^2 \rho_{\rm p} \Delta {\rm v}
\simeq \pi b_{\rm set}^2 \frac{\Sigma_{\rm p}}{\sqrt{2\pi} h_{\rm p}} \Delta {\rm v}
\nonumber \\
 & \simeq \frac{r}{\sqrt{2\pi} h_{\rm p}} \eta^{-1} \frac{M_{\rm pl}}{M_\ast} \dot{M}_{\rm peb}
\equiv f_{\rm flt} \dot{M}_{\rm peb},
\label{eq:M_pl_dot}
\end{align}
where $\rho_{\rm p}$ is the spatial mass density of the pebbles
and we used Eqs.~(\ref{eq:mdot_peb}) and (\ref{eq:pib2}).
The parameter $f_{\rm flt}$ expresses the mass fraction of the accretion flow onto the planet
in the pebble flux through the disk before the accretion, 
which is called a "filtering factor." 
Note that $f_{\rm flt}$ does not directly depend on $\Delta {\rm v}$, $b_{\rm set}$, and ${\rm St}$.

When $b_{\rm set} > h_{\rm p}$, the accretion is 2D and
$\pi b_{\rm set}^2 \rho_{\rm p}$ 
in Eq.~(\ref{eq:M_pl_dot}) is replaced by $2 b_{\rm set} \Sigma_{\rm p}$.
Accordingly, a complete filtering factor is
\begin{align}
f_{\rm flt} = \min \left(\frac{2r}{\pi b_{\rm set}}, \frac{r}{\sqrt{2\pi} h_{\rm p}} \right) \eta^{-1} \frac{M_{\rm pl}}{M_\ast}.
\label{eq:acc-rate}
\end{align}
While $f_{\rm flt}$ depends on $b_{\rm set}$ in 2D mode,
it still does not depend directly on ${\rm St}$ and $\Delta {\rm v}$.
In our simulations, we set the planetary embryos
 with the masses and semimajor axises identical to 
Venus, Earth and Mars (eccentricities are set to be zero).
In these cases, the relative velocity is given by $\Delta {\rm v} = (3/2) b_{\rm set} \Omega$
("Hill regime").
Substituting this into Eq.~(\ref{eq:pib2}), 
\begin{align}
b_{\rm set} \approx 2 {\rm St}^{1/3} \left( \frac{M_{\rm pl}}{3 M_{\ast}} \right)^{1/3} r.
\label{eq:b_set_hill}
\end{align}
In the numerical simulations, we used a more general formula including "Bondi regime" and
the cut-off parameter ($\exp[-({\rm St} / 2)^{0.65}]$) for large St cases
\citep{OrmelKobayashi2012}. 
Note that the dependence on $\Delta {\rm v}$ cancels in the accretion rate both in 2D and 3D cases
and $\dot{M}_{\rm pl}$ does not directly depend on $b_{\rm set}$ and ${\rm St}$ in the 3D case.
For a small planet, the accretion is in 3D mode. 
The accretion becomes 2D mode when
\begin{align}
M_{\rm pl} > M_{\rm 2D3D} & \simeq 
3\left(\frac{2}{\pi}\right)^{3/2} {\rm St}^{-1} 
\left(\frac{\rm St}{\alpha}\right)^{-3/2}
\left(\frac{h_{g}}{r}\right)^{3} M_* \nonumber \\
 & \simeq 1.7 \left( \frac{M_{\ast}}{M_{\odot}} \right)^{-5/7} \left( \frac{\rm St}{0.1} \right)^{-5/2} \left( \frac{\alpha}{10^{-2}} \right)^{3/2} \left( \frac{r}{1 {\rm au}} \right)^{6/7} M_{\oplus}.
\label{eq:M_2D3D}
\end{align}
The filtering factor is given by
\begin{equation}
f_{\rm flt} \simeq
\left\{
\begin{array}{l} 
\displaystyle 0.017\left( \frac{M_{\ast}}{M_{\odot}} \right)^{-1} \left(\frac{\alpha}{10^{-2}} \right)^{-1/2} 
\left(\frac{h_{\rm g}/r}{0.02}\right)^{-1} 
\left(\frac{\rm St}{0.1}\right)^{1/2} 
\left(\frac{M_{\rm pl}}{0.1 M_\oplus}\right)
\left(\frac{r}{1 {\rm au}} \right)^{-1/2}  \\
\hspace*{5cm} [M_{\rm pl} < M_{\rm 2D3D}; {\rm 3D}] \\
 \\
\displaystyle 0.040 \left( \frac{M_{\ast}}{M_{\odot}} \right)^{-2/3} 
\left(\frac{\rm St}{0.1}\right)^{-1/3} 
\left(\frac{M_{\rm pl}}{0.1 M_\oplus}\right)^{2/3} 
\left(\frac{r}{1 {\rm au}} \right)^{-1/2}  \\
\hspace*{5cm} [M_{\rm pl} > M_{\rm 2D3D}; {\rm 2D}],
\end{array}
\right.
\label{eq:fflt}
\end{equation}
where we used Eq.~(\ref{eq:eta}).

We take into account the decrease in dust surface density and pebble flux 
due to the pebble accretion onto the planets.
After the snowline passage, Eq.~(\ref{eq:dust_menpd}) is rewritten 
in the grid where the planets exist, as
\begin{align}
\frac{\partial \Sigma_{\rm p}}{\partial t} + \frac{1}{r} \frac{\partial}{\partial r} (r {\rm v}_{\rm r,d} \Sigma_{\rm p}) + \frac{\dot{M}_{\rm pl}}{2 \pi r \Delta r} = 0,
\label{eq:dust_acc}
\end{align}
where $\Delta r$ is the radial grid size. 
After the icy pebble accretion onto the planet starts, the pebble mass flux decreases discontinuously at the planet orbits,
according to the accretion onto the planets.

\subsection{Simulation parameters}

The parameters for our simulations are the initial disk characteristic size $r_{\rm d,0}$,
 the diffusion timescale of the disk $t_{\rm diff}$, the initial disk gas accretion rate $\dot{M}_{\rm g,0}$, 
 and the photo-evaporation rate $\dot{M}_{\rm pe}$.
 As we explained in section 2.1, the other disk parameters are calculated by these parameters.
We perform simulations with

1) $\dot{M}_{\rm g,0} = 3\times 10^{-8}, 10^{-7} M_\odot/{\rm y}$,

2) $\dot{M}_{\rm pe} = 10^{-9}, 10^{-8} M_\odot/{\rm y}$,

3) $r_{\rm d,0}=30, 100, 300$ au,

4) $t_{\rm diff} = 10^6, 3 \times 10^6$ y.

The initial total gas disk mass is given by Eq.~(\ref{eq:Sigma_g2}) as
\begin{align}
M_{\rm g,0} & = 2 t_{\rm diff} \,(\dot{M}_{\rm g,0} -  \dot{M}_{\rm pe}) \nonumber\\
 & = 0.06 \left(\frac{t_{\rm diff}}{10^6{\rm y}}\right) 
 \left(\frac{\dot{M}_{\rm g,0} -  \dot{M}_{\rm pe}}{3\times 10^{-8} M_\odot/{\rm y}}\right) M_\odot, 
\end{align}
which ranges from $0.06 M_\odot$ to $0.6 M_\odot$ with the above parameters.
Because we start with relatively large $\dot{M}_{\rm g,0}$, the initial disk mass is relatively large. 
As we will show in section 4, initially massive disks
tend to produce dry planets, while small mass disks tend to produce water-rich planets.  

The initial dust surface density distribution is simply given by $0.01 \, \Sigma_{\rm g,0}$.
With the gas disk evolution, we simultaneously calculate the pebble growth from the dust disk
and its migration.
The dust disk is depleted by the formation and 
radial drift of pebbles.

We set Venus, Earth and Mars analogues
in circular orbits 
with the same masses and the semimajor axises as the current Venus, Earth and Mars in Solar system.
For the results presented here, we assume $M_* = M_\odot$ and $L_* = L_\odot$,
while we retain the dependence on $M_*$ and $L_*$ in the equations. 
For other-mass stars, the dependence of 
$\dot{M}_{\rm g,0}, \dot{M}_{\rm pe}, r_{\rm d,0}$ and $t_{\rm diff}$
on the stellar mass also have to be considered.

\section{Numerical Results}

\subsection{Icy grain/pebble evolution}

We calculate the growth and radial drift of icy grains and pebbles
following the method by \cite{Sato2016}, which is described in section 2.2.
The evolution of icy particles (grains and pebbles)
found by simulations of \citet{Sato2016} is summarized as follows:
\begin{enumerate}
\item 
The growth timescale of a particle with mass $m_{\rm p}$
is well approximated by the simple formula with
$\Delta {\rm v_{pp}} \simeq \Delta {\rm v}_{\rm t}$ in Epstein regime
(also see \citealt{TakeuchiLin2005}; \citealt{Brauer2008}) as
 \begin{equation}
 t_{\rm grow} = \frac{m_{\rm p}}{dm_{\rm p}/dt}
 \simeq \frac{4}{\sqrt{3\pi}} Z_0^{-1} \Omega^{-1},
 \label{eq:tgrow}
 \end{equation}
 where $Z_0$ is the initial particle-to-gas ratio in the disk.
 The timescale of growth from $\mu$m-size grains to pebbles 
 by several orders of magnitude in radius is
 \begin{align}
 t_{\rm grow,peb} & \sim 10 \, t_{\rm grow} \nonumber\\
  & \simeq 2 \times 10^5 \left(\frac{Z_0}{10^{-2}}\right)^{-1} 
 \left(\frac{r}{100\,{\rm au}}\right)^{3/2}
 \left(\frac{M_\ast}{M_\odot}\right)^{-1/2}\; {\rm y}.
 \label{eq:tgrowp}
 \end{align}
 
\item 
The drift timescale becomes shorter as $m_{\rm p}$ 
(equivalently, Stokes number St) increases (Eq.~\ref{eq:tmig}),
while the growth timescale is independent of $m_{\rm p}$ (Eq.~\ref{eq:tgrow}). 
When $t_{\rm drift}$ becomes smaller than $t_{\rm grow}$,
the particle drift effectively starts and $\Sigma_{\rm p}$ starts being sculpted.
From Eqs.~(\ref{eq:tmig}) and (\ref{eq:tgrow}) 
with $Z_0$ replaced by the particle-to-gas ratio of drifting pebbles ($Z$),
the equilibrium Stokes number of drifting pebbles is
\begin{equation}
{\rm St} \sim \frac{\sqrt{3\pi}}{8 \, \eta} Z
\sim 0.03 \left(\frac{Z_0}{10^{-2}}\right) \left(\frac{r}{100 \,{\rm au}}\right)^{-1/2},
\label{eq:eqSt}
\end{equation}
where we used a typical value of the solid-to-gas ratio of
migrating pebble, $Z \sim 0.1 Z_0$  \citep{Ida-et2016}, where we adopt $Z_0$
is the initial particle-to-gas ratio and adopt $Z_0=0.01$ for a nominal value.

\item 
The sensitive $r$-dependence of $t_{\rm grow,peb}$ 
results in ''an inside-out formation" of pebbles; formation is earlier in inner region and
the formation front migrates outward
\citep[also see][]{Brauer2008,Okuzumi2012,Birnstiel2012,LJ2014}.
After the pebble formation front reaches the disk outer edge, 
$\Sigma_{\rm p}$ rapidly decays uniformly in the disk,
because the supply from further outer regions to the outermost region is limited.
\end{enumerate}

As we will show below,
the time ($t_{\rm pff}$) at which 
the pebble formation front reaches the disk outer edge 
is a very important parameter.
The timescale for the pebble formation front to
reach the radius of $r_{\rm d,0}$ is
given with Eq.~(\ref{eq:tgrowp}) by
\begin{equation}
 t_{\rm pff} \sim 2 \times 10^5 \left(\frac{Z_0}{10^{-2}}\right)^{-1} 
 \left(\frac{r_{\rm d,0}}{100\,{\rm au}}\right)^{3/2}
 \left(\frac{M_\ast}{M_\odot}\right)^{-1/2}\; {\rm y}.
 \label{eq:tpff0}
 \end{equation}
In other words, the pebble formation front radius is given as a function of time ($t$) by
\begin{equation}
 r_{\rm pff} \sim 100 \left(\frac{t}{2 \times 10^5{\rm y}}\right)^{2/3}
  \left(\frac{Z_0}{10^{-2}}\right)^{2/3} 
 \left(\frac{M_\ast}{M_\odot}\right)^{1/3}\; {\rm au}.
 \label{eq:tpff}
 \end{equation}
The timescale $t_{\rm pff}$ is determined only by $r_{\rm d,0}$ in the disk parameters.
It is independent of the other disk parameters such as $t_{\rm diff}$ and $\dot{M}_{\rm g,0}$.

\begin{figure*}\begin{center} 
\includegraphics[width=80mm]{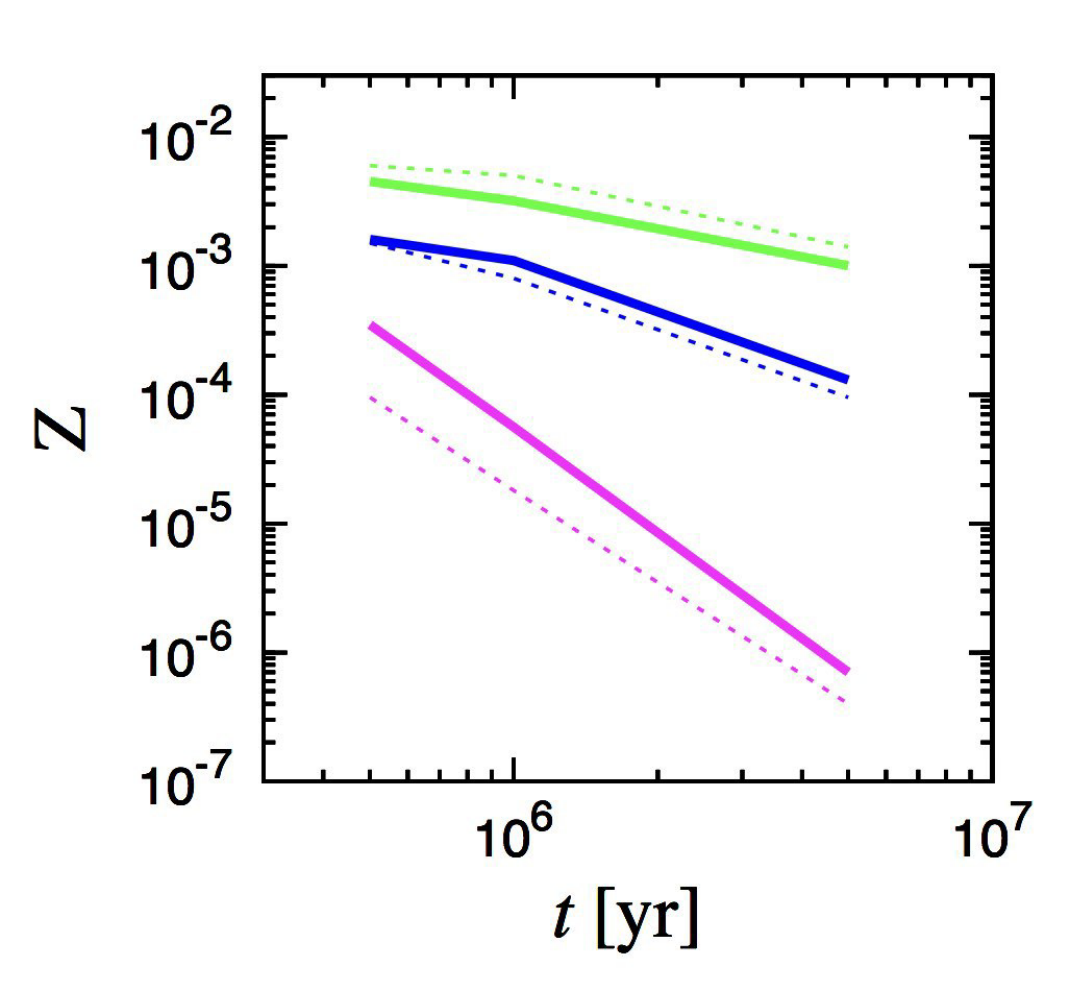}
\end{center}
\caption{The time evolution of 
$Z = \Sigma_{\rm p}/\Sigma_{\rm g}$ obtained by the numerical simulation
with $\dot{M}_{\rm g,0}=10^{-7}M_\odot/{\rm y}$, 
$\dot{M}_{\rm pe}=10^{-9}M_\odot/{\rm y}$, and $t_{\rm diff}=10^6{\rm y}$.
The magenta, blue, and green solid lines are
$Z$ obtained at $\rd0$ by the numerical results of with $\rd0=30, 100$ and 300 au. 
The dashed lines are gradients for
individual cases given by Eqs.~(\ref{eq:Sigma_decay}) and (\ref{eq:gamma2}),
where the absolute values of $Z$ are arbitrary.
}\label{fig:sigma_evol}
\end{figure*}

Once the formation front reaches $r_{\rm d,0}$, 
the supply of solid materials from further outer regions is limited.
Accordingly, the pebble formation and drift from there 
results in the decay of $\Sigma_{\rm p}$ near $r_{\rm d,0}$.
The solid surface density at $r < r_{\rm d,0}$ also decays, 
because it is contributed from drifting pebbles formed near the formation front. 
Thus, the decay rate of $\Sigma_{\rm p}$ for $t > t_{\rm pff}$ is regulated by
the pebble growth near $r_{\rm d,0}$, and 
is approximated as $\partial \Sigma_{\rm p}/\partial t \sim - \Sigma_{\rm p}/t_{\rm pff} 
\simeq - (\sqrt{3\pi}/40)(\Sigma_{\rm p}^2/\Sigma_{\rm g})\Omega$.
From this relation, $Z = \Sigma_{\rm p}/\Sigma_{\rm g} \propto t^{-1}$ at $t > t_{\rm pff}$,
which is approximated as 
$Z  \propto (1 + t/t_{\rm pff})^{-1}$.
However,  the numerical results show a faster decrease due to the effect of finite $\rd0$
\citep{Sato2016}.
The numerically obtained time evolution of $Z$ at $r=\rd0$ with $\rd0=30, 100$ and 300 au is shown in Fig.~\ref{fig:sigma_evol}.
 For $\rd0=30, 100$ and 300 au,
$t_{\rm pff} \simeq 3.3 \times 10^4, 2.0 \times 10^5$ and $1 \times 10^6$ y,
respectively (Eq.~\ref{eq:tpff}).
We fit the numerical results as
\begin{align}
Z & \propto (1 + t/t_{\rm pff})^{-\gamma},
\label{eq:Sigma_decay} 
\end{align}
where 
\begin{align}
& \gamma = 1 + \gamma_2 (300\,{\rm au}/\rd0), \nonumber
\\
& \gamma_2 \sim 0.15.
\label{eq:gamma2}
\end{align}
Because $t_{\rm pff} \ga t_{\rm diff}$ and the finite $r_{\rm d,0}$ does not affect
at $r \ga 300\,{\rm au}$ (Eq.~\ref{eq:tpff}),
the $\rd0$-dependence as a factor of $(300\,{\rm au}/\rd0)$ is reasonable.
Although the value of $\gamma_2$ may include an uncertainty
due to the disk model and the single-size approximation,
we will show in section 5 that
the predicted function of the water fraction depends only weakly on $\gamma_2$.

Note that we assume that collisions between icy pebbles always result in sticking and
the pebble size is determined by the drift limit.
If the pebble growth is limited by bouncing collisions or collisional fragmentation,
St is determined by the threshold velocity for bouncing or fragmentation, which is lower than
the value in Eq.~(\ref{eq:eqSt}), and Equation~(\ref{eq:tpff}) depends on the threshold St.
It is a widely accepted idea that pebbles made of H$_2$O ice grains have a high sticking threshold velocity \citep{Wada09,GundlachBlum15}, but this will not be the case if the grains are mantled by poorly sticky CO$_2$ \citep{Musiolik16}. 
Recent studies have found that this CO$_2$-induced fragmentation can have important implications for the growth of pebble-accreting protoplanets \citep{Johansen2018} and for the observational appearance of protoplanetary disks (Okuzumi and Tazaki 2019, submitted).
How the bouncing and fragmentation barriers affect the water delivery to rocky planets will be studied in future work.

\subsection{Water fraction}

With the simulated pebble mass flux $\dot{M}_{\rm peb}$,
we calculated the growth rate of a planet due to icy pebble accretion ($\dot{M}_{\rm pl}$).
The filtering factor is defined by $ f_{\rm flt} = \dot{M}_{\rm pl}/\dot{M}_{\rm peb}$.
When the snowline passes each planetary orbit,
we switch on the icy pebble accretion onto the planet.
We assume that the 1:1 ratio of rocky to icy fraction of icy pebbles.
We set the rocky planets with the Venus, Earth, and Mars masses
when the snowline passes 0.72 au, 1.0 au and 1.52 au, respectively.
As we will show, the water fraction of final planets is insensitive to $M_{\rm pl,0}$.

After the snowline passes a planetary orbit,
icy pebble accretion starts.
When the cumulative accreted mass by icy pebbles
is $\Delta M_{\rm pl}$,
the total ice mass in the planet is $(1/2) \Delta M_{\rm pl}$.
The water fraction of the planet at the mass $M_{\rm pl}$ 
is given by 
\begin{equation}
f_{\rm water} = 
\frac{(1/2) \Delta M_{\rm pl}}{M_{\rm pl}} =
\frac{1}{2} \frac{\Delta M_{\rm pl}}{M_{\rm pl,0}+\Delta M_{\rm pl}},
\label{eq:fwater}
\end{equation}
where $M_{\rm pl,0}$ is the planetary mass at the snowline passage,
$M_{\rm pl} = M_{\rm pl,0}+\Delta M_{\rm pl}$.
If $\Delta M_{\rm pl}$ becomes much larger than $M_{\rm pl,0}$, 
the water fraction saturates to $f_{\rm water} \simeq 1/2$.

We repeat the simulations with different disk parameters,
$t_{\rm diff}$, $\dot{M}_{\rm g,0}$, 
$r_{\rm d,0}$, and $\dot{M}_{\rm pe}$,
to investigate how the water fraction of the final planets depends on these parameters
and which values of the parameters produce the water fraction consistent with the terrestrial planets in the Solar system. 

\begin{figure*}
\vspace*{-1cm}
\begin{minipage}{0.33\hsize} \begin{center} \includegraphics[width=65mm]{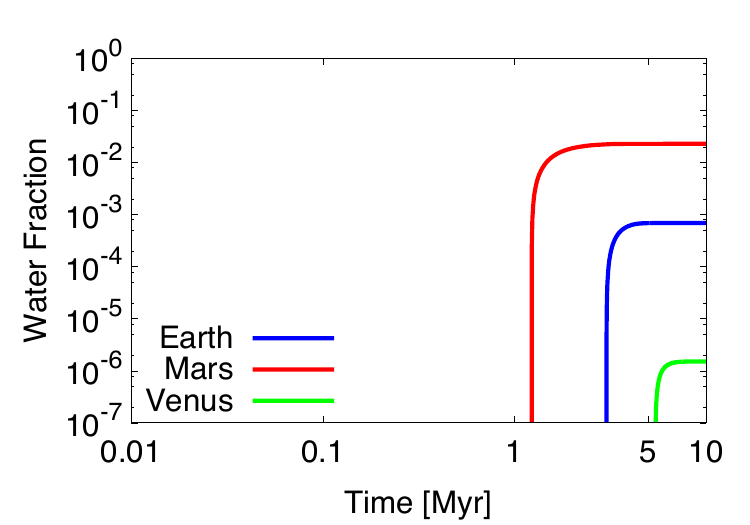} \end{center} \end{minipage}
\begin{minipage}{0.33\hsize}\begin{center} \includegraphics[width=65mm]{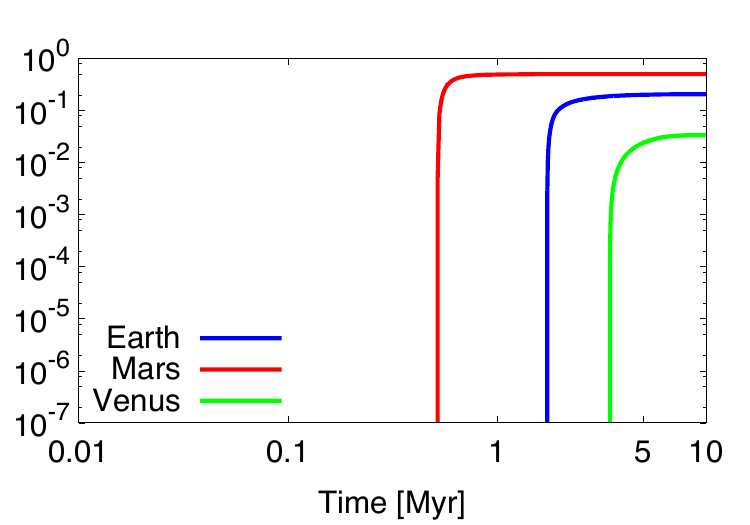} \end{center}\end{minipage}
\begin{minipage}{0.33\hsize}\begin{center} \includegraphics[width=65mm]{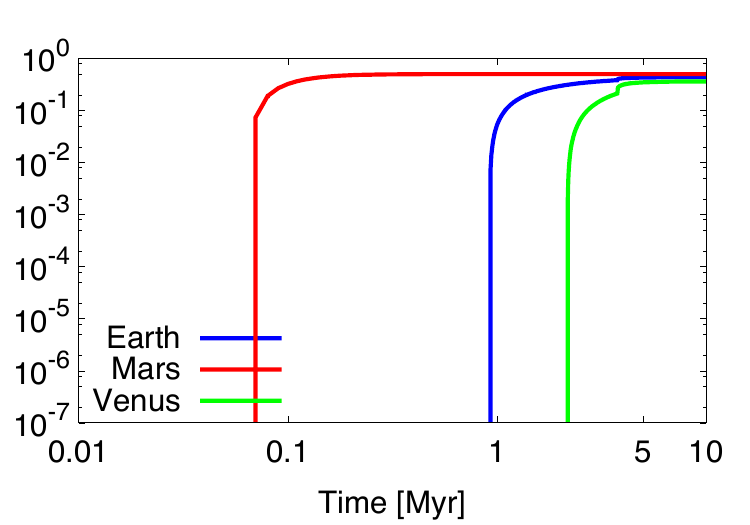}\end{center} \end{minipage}\vspace{20pt}
\caption{
The time evolution of water fraction
for the models with $t_{\rm diff} = 10^6$y, $\dot{M}_{\rm g,0} = 10^{-7} M_{\odot} / {\rm y}$ and $\dot{M}_{\rm pe} = 10^{-9} M_{\odot} / {\rm y}$. 
The left, center and right panel show the results of $r_{\rm d.0} = 30 {\rm au}, 100{\rm au}$, and $300 {\rm au}$, respectively.
The blue, red, and green lines represent the Earth, Mars and Venus analogues, respectively. 
}
\label{fig:water-rout}
\end{figure*}

Figure~\ref{fig:water-rout} shows the time evolution of water fraction for the models with $t_{\rm diff} = 10^6$y, 
$\dot{M}_{\rm g,0} = 10^{-7} M_{\odot} / {\rm y}$, and $\dot{M}_{\rm pe} = 10^{-9} M_{\odot} / {\rm y}$.
The left, middle and right panels show the results of $r_{\rm d,0} = 30$, 100, 
and $300\,{\rm au}$, respectively.
We first explain general evolution pattern of the water fraction.
In all cases, the water fraction rapidly increases once the snowline passes the planetary orbit and the icy pebble accretion starts.
It is saturated to its asymptotic value, even sufficiently before completion of disk gas depletion.
From Eqs.~(\ref{eq:r_snow_vis}) and (\ref{eq:alpha}),
\begin{align}
r_{\rm snow,vis} \simeq & \; 2.1\, \left(\frac{M_*}{M_\odot}\right)^{1/3} 
\nonumber \\
 & \times \left(\frac{r_{\rm d,0}}{100\,{\rm au}}\right)^{-2/9} 
\left(\frac{t_{\rm diff}}{10^6\,{\rm y}}\right)^{2/9} 
\left(\frac{\dot{M}_{\rm g}}{10^{-7}M_\odot/{\rm y}}\right)^{4/9} {\rm au}.
\label{eq:r_snow2}
\end{align}
At $t = 0$ with $\dot{M}_{\rm g,0} = 10^{-7} M_{\odot} / {\rm y}$,
the snowline is located outside Mars' orbit.
As shown in Fig.~\ref{fig:lines}, 
the snowline migrates inward and passes through the planetary orbits one after another
as the disk accretion rate $\dot{M}_{\rm g}$
decreases with time as Eq.~(\ref{eq:flux2}).
The snowline passage always occurs in the order of Mars at 1.52 au, Earth at 1.0 au,
and Venus at 0.72 au.
Because the initial $r_{\rm snow,vis}$ is closer to the planetary orbits for larger $r_{\rm d,0}$,
the snowline passage is earlier for larger $r_{\rm d,0}$, 
as Figure~\ref{fig:water-rout} shows.
The water fraction is saturated when $\Sigma_{\rm p}$ and $\dot{M}_{\rm peb}$
significantly decay.
The rapid decay starts at $t \sim t_{\rm pff}$ and  
Eq.~(\ref{eq:tgrowp}) shows that $t_{\rm pff}$ is 
as short as $\sim {\rm a \; few} \times 10^5$ y for $r_{\rm d,0} = 100$ au.
Even for $r_{\rm d,0} = 300$ au, the decay starts 
at $t_{\rm pff} \simeq 10^6$ y before disk depletion.
Therefore, the evolution of the water fraction evolution is mainly reduced
by the consumption of icy dust grain reservoir rather than by the disk gas depletion.

Figure~\ref{fig:water-rout} shows a clear trend that 
the final water fraction is lower for smaller $r_{\rm d,0}$.
This trend is explained by a comparison between
the snowline passage time $t_{\rm snow}$ and $t_{\rm pff}$, as follows.
The water fraction due to the pebble accretion rapidly increases 
until $\Sigma_{\rm p}$ decays by more than an order of magnitude,
which corresponds to, say, $t > 10\, t_{\rm pff}$ (Eq.~\ref{eq:Sigma_decay}). 
If $t_{\rm snow} > 10 \, t_{\rm pff}$,
$f_{\rm water}$ can be $\ll 1$.
While $t_{\rm pff}$ is smaller for a smaller $r_{\rm d,0}$
(Eq.~\ref{eq:tgrowp}), 
$t_{\rm snow}$ is larger (Eq.~\ref{eq:r_snow2}).
The latter implies that the disk is warmer for a smaller $r_{\rm d,0}$.
The viscous heating increases as 
the blanketing effect by optical depth ($\propto \Sigma_{\rm g}$) increases.
In Fig.~\ref{fig:water-rout}, $\dot{M}_{\rm g,0}$ and $t_{\rm diff}$ are fixed.
Smaller $r_{\rm d,0}$ means smaller $\alpha$ (Eq.~\ref{eq:alpha_derive})
and larger $\Sigma_{\rm g}$ (Eq.~\ref{eq:Sigma_g2}), resulting in a warmer disk. 
Thereby, the final water fraction is lower for smaller $r_{\rm d,0}$.

We examine the condition of $t_{\rm snow}/t_{\rm pff} > 10$
or $ < 10$ in more details.
In the case of $r_{\rm d,0}=30$ au,  
$t_{\rm pff} \simeq 4 \times 10^4$ y (Eq.~\ref{eq:tgrowp}) 
and $t_{\rm snow}$ is identified by the timing at which $f_{\rm water}$
starts rapid increase, which is $\simeq 1 \times 10^6$ y, $2.5 \times 10^6$ y and
$5 \times 10^6$ y for the Mars, Earth and Venus analogues, respectively  
(the left panel of Figure~\ref{fig:water-rout}).
Because $t_{\rm snow} > 10\, t_{\rm pff}$ in this case, $f_{\rm water} \sim 10^{-2}$
even for the outermost Mars analogue.
For the Earth and Venus analogues, $f_{\rm water}$ is further smaller.
In the case of $r_{\rm d,0}=300$ au,  
$t_{\rm pff} \simeq 1 \times 10^6$ y (Eq.~\ref{eq:tgrowp}) 
and $t_{\rm snow} \simeq 1 \times 10^5$ y, $1 \times 10^6$ y and
$2 \times 10^6$ y for Mars, Earth, and Venus analogues, respectively (the right panel of Figure~\ref{fig:water-rout}).
Because $t_{\rm snow} < 10\, t_{\rm pff}$, 
$f_{\rm water} \sim 1/2$ for all of Mars, Earth and Venus.
In the middle panel of Figure~\ref{fig:water-rout}, $r_{\rm d,0}=100$ au
and $t_{\rm pff} \simeq 2 \times 10^5$ y (Eq.~\ref{eq:tgrowp}).
In this case, $t_{\rm snow} < 10 \, t_{\rm pff}$ and $f_{\rm water} \ll 1$ 
for the Venus analogue, while
$t_{\rm snow} > 10\, t_{\rm pff}$ and $f_{\rm water} \sim 1/2$ for the Mars analogue.

\begin{figure*}
\vspace*{-1cm}
\begin{minipage}{0.5\hsize} \begin{center}  \includegraphics[width=65mm]{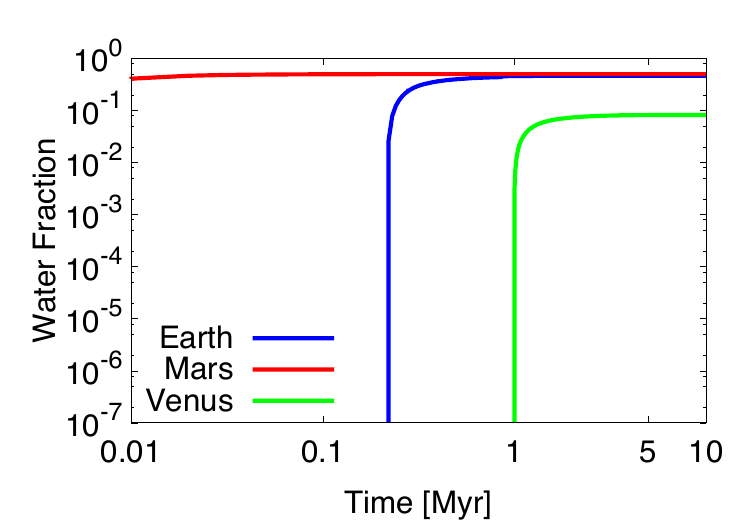} \end{center} \end{minipage}
\begin{minipage}{0.5\hsize}\begin{center} \includegraphics[width=65mm]{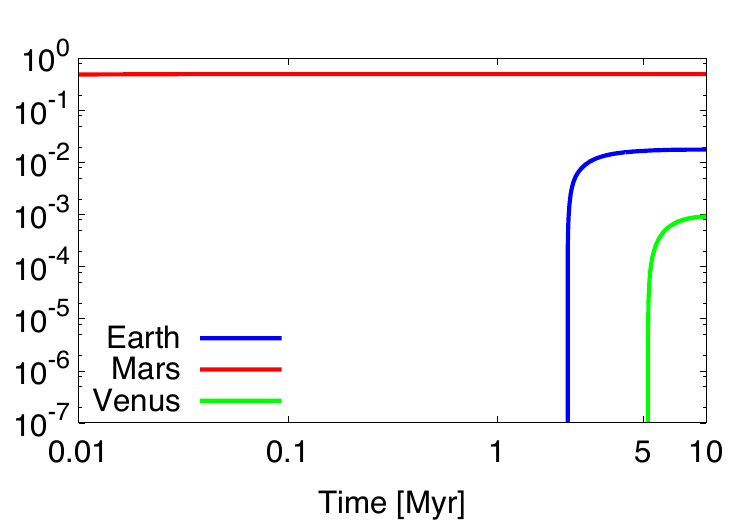}\end{center}\end{minipage} 
\vspace{20pt}
\caption{
The time evolution of water fraction for the models with $r_{\rm d.0} = 100$au, $\dot{M}_{\rm g,0} = 3 \times 10^{-8} M_{\odot} / {\rm y}$ and $\dot{M}_{\rm pe} = 10^{-9} M_{\odot} / {\rm y}$. The left and right panels show the results of $t_{\rm diff} = 10^6 {\rm y}$, and $3 \times 10^6 {\rm y}$, respectively.
}
\label{fig:water-tdiff}
\end{figure*}

The condition of 
$t_{\rm snow}/t_{\rm pff} > 10$ or $ < 10$
also explains other results
of the final water fraction of planets,
because this condition represents 
how much icy materials remain at the snowline passage.
(In section 5, we will revisit this condition.)
Figure~\ref{fig:water-tdiff} is the time evolution of water fraction 
of $t_{\rm diff} = 10^6 {\rm y}$ (the left panel) and
$t_{\rm diff} = 3 \times 10^6 {\rm y}$ (the right panel), respectively.
The other disk parameters are the same.
For both cases, $10 \, t_{\rm pff} \simeq 2 \times 10^6$ y (Eq.~\ref{eq:tgrowp}).
The snowline is already inside Mars' orbit from the beginning ($t = 0$) of
the calculations, that is, $t_{\rm snow}=0$, 
which results in $f_{\rm water} \simeq 1/2$ for the Mars analogue.
The snowline passage time $t_{\rm snow}$ is
smaller than $10 \, t_{\rm pff}$ only for the Venus analogue
in the case of $t_{\rm diff} = 10^6 {\rm y}$ (the left panel),
and for both the Earth and Venus analogues
in the case of $t_{\rm diff} = 3 \times 10^6 {\rm y}$ (the right panel).
This explains the results in Fig.~\ref{fig:water-tdiff}.

\begin{figure*}
\vspace*{-1cm}
\begin{minipage}{0.5\hsize}\begin{center}  \includegraphics[width=65mm]{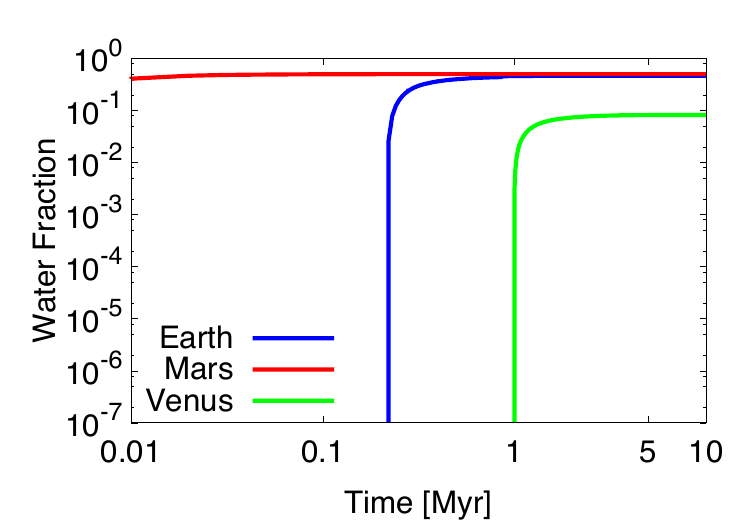}\end{center}\end{minipage}
\begin{minipage}{0.5\hsize}\begin{center} \includegraphics[width=65mm]{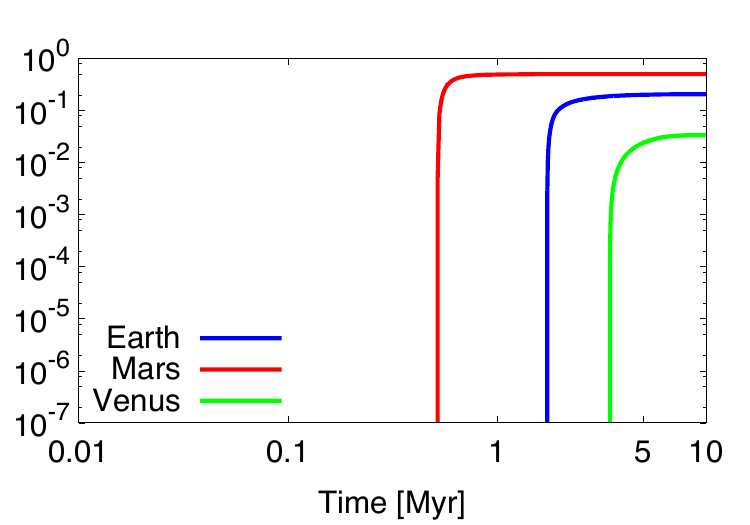}\end{center}\end{minipage}
\vspace{20pt}
\caption{
The time evolution of water fraction for the models with $r_{\rm d.0} = 100$au, $t_{\rm diff} = 10^6$y and $\dot{M}_{\rm pe} = 10^{-9} M_{\odot} / {\rm y}$. The left and right panels show the results of, $\dot{M}_{\rm g,0} = 3 \times 10^{-8}M_{\odot} / {\rm y}$ and $10^{-7} M_{\odot} / {\rm y}$, respectively.
}
\label{fig:water-flux}
\end{figure*}

\begin{figure*}
\vspace*{-1cm}
\begin{minipage}{0.33\hsize}\begin{center}  \includegraphics[width=65mm]{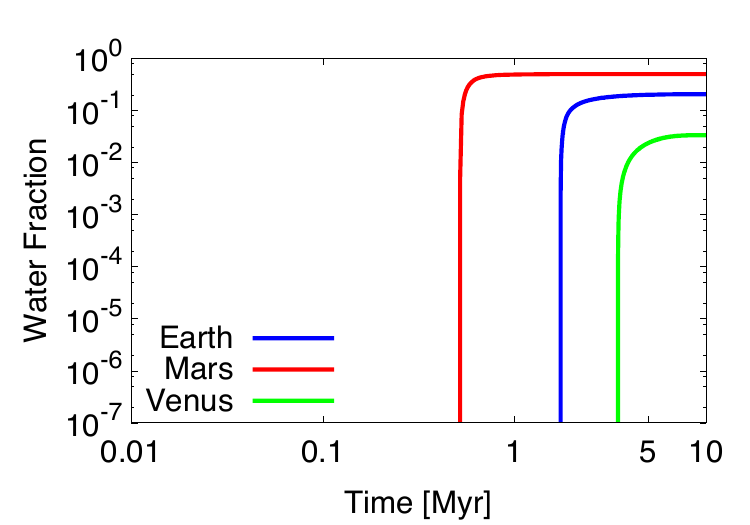}\end{center}\end{minipage}
\begin{minipage}{0.33\hsize}\begin{center} \includegraphics[width=65mm]{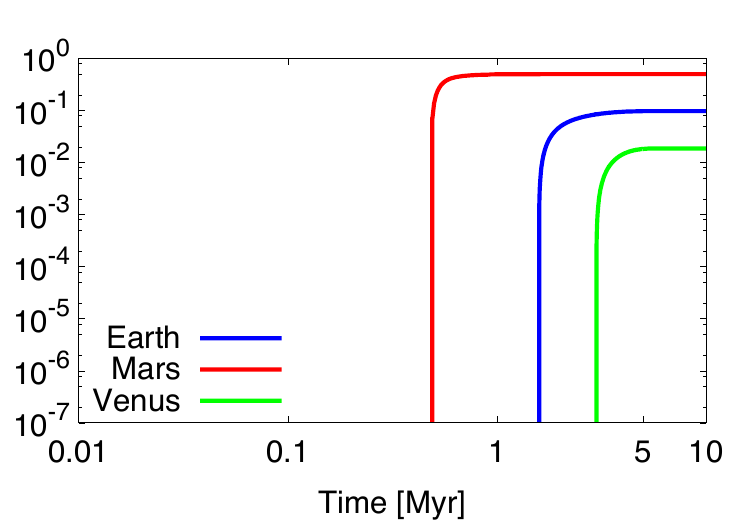}\end{center} \end{minipage}
\begin{minipage}{0.33\hsize}\begin{center} \includegraphics[width=65mm]{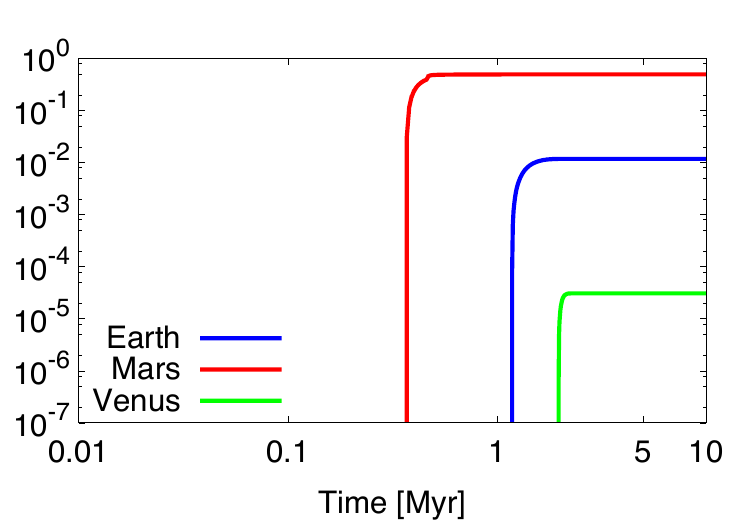}\end{center}\end{minipage}
\vspace{20pt}
\caption{
The time evolution of water fraction for the models with $r_{\rm d.0} = 100$au, $t_{\rm diff} = 10^6$y and $\dot{M}_{\rm ini} = 10^{-7} M_{\odot} / {\rm y}$. The left, center and right panels show the results of, $\dot{M}_{\rm pe} = 10^{-9}M_{\odot} / {\rm y}, 3 \times 10^{-9}M_{\odot} / {\rm y}$, and $10^{-8} M_{\odot} / {\rm y}$, respectively.
}
\label{fig:water-fluxpe}
\end{figure*}

Figure~\ref{fig:water-flux} shows the results of
$\dot{M}_{\rm g,0} = 3 \times 10^{-8}M_{\odot} / {\rm y}$ (left) and $10^{-7} M_{\odot} / {\rm y}$ (right)
with $r_{\rm d,0} = 100$ au, $t_{\rm diff} = 10^6$y, 
and $\dot{M}_{\rm pe} = 10^{-9} M_{\odot} / {\rm y}$.
Again, $10 \, t_{\rm pff} \simeq 2 \times 10^6$ y for both cases.
Smaller $\dot{M}_{\rm g,0}$ with the fixed $t_{\rm diff} = 10^6 {\rm y}$ means
earlier passage of the snowline, resulting in a lower water fraction.
Figure~\ref{fig:water-fluxpe} shows the results of $\dot{M}_{\rm pe} = 10^{-9}M_{\odot} / {\rm y}$
(left) and $\dot{M}_{\rm pe} = 3 \times 10^{-9}M_{\odot} / {\rm y}$ (middle) and $\dot{M}_{\rm pe}= 10^{-8} M_{\odot} / {\rm y}$.
With larger $\dot{M}_{\rm pe}$, the outer disk region is truncated at shorter radius.
The disk truncation reduces the reservoir of icy materials,
resulting in a lower water fraction, 
for the planets (the Earth and Venus analogues) with 
$t_{\rm snow} \ga 10 \, t_{\rm pff} \simeq 2 \times 10^6$ y.

In section 4, we will derive a semi-analytical formula to predict the water fraction
of planets after disk depletion.
We will show that the total mass ($M_{\rm res}$) 
of the icy dust materials preserved in the disk 
determines the final water fraction, because they eventually 
drift to the inner regions and pass the planetary orbits\footnote{
The grains at $r > 300-500 \, {\rm au}$ would not undergo radial drift sufficiently,
because their growth timesacle is longer than $t_{\rm diff}$.
We here consider disks with $r_{\rm d,0} \le 300 \,{\rm au}$.}.  
Since our gas disk model is analytical, we can analytically
evaluate $t_{\rm snow}$.
We already know the analytical expression of $t_{\rm pff}$ given by Eq.~(\ref{eq:tpff})
is a good approximation.
We also semi-analytically derived how the icy grain surface density evolves
as Eq.~(\ref{eq:Sigma_decay}). 
Synthesis of these results enables us to 
derive the semi-analytical formula for $f_{\rm water}$.

Finally we point out that $f_{\rm water}$ is higher in the order of Mars,
Earth, and Venus analogues according to the snowline passage timing,
except for the fully saturated cases where $f_{\rm water} \sim 1/2$ for all the planets.
We will show that the final water fraction is insensitive to the embryo mass.
The timing is the most important factor for $f_{\rm water}$.

In the next section, we show that the derived semi-analytical expression of $f_{\rm water}$
reproduces the numerical results.
Using the analytical expression, we will clarify the dependence
of the final water fraction on the disk parameters and 
pebble accretion parameters such as the initial planetary mass and Stokes number of
accreting pebbles in section 5.
We will also survey the disk parameter range
that may reproduce $f_{\rm water}$ comparable to that of the terrestrial planets in our Solar system.

\section{Analytical formula for planetary water fraction}

The planetary water fraction is calculated by
estimating the cumulative mass of accreted icy pebbles $\Delta M_{\rm pl}$
(Eq.~\ref{eq:fwater}).
We can simply estimate it as $\Delta M_{\rm pl} \simeq f_{\rm flt}(M_{\rm pl}) \cdot M_{\rm res}$, where 
$M_{\rm res}$ is the total icy dust mass preserved in outer disk regions 
at the snowline passage, because 
the pebble flux integrated from the snowline passage time $t_{\rm snow}$
to infinity (effectively, to $t_{\rm diff}$) must be similar to $M_{\rm res}$.

\begin{figure*}\begin{center} 
\includegraphics[width=70mm]{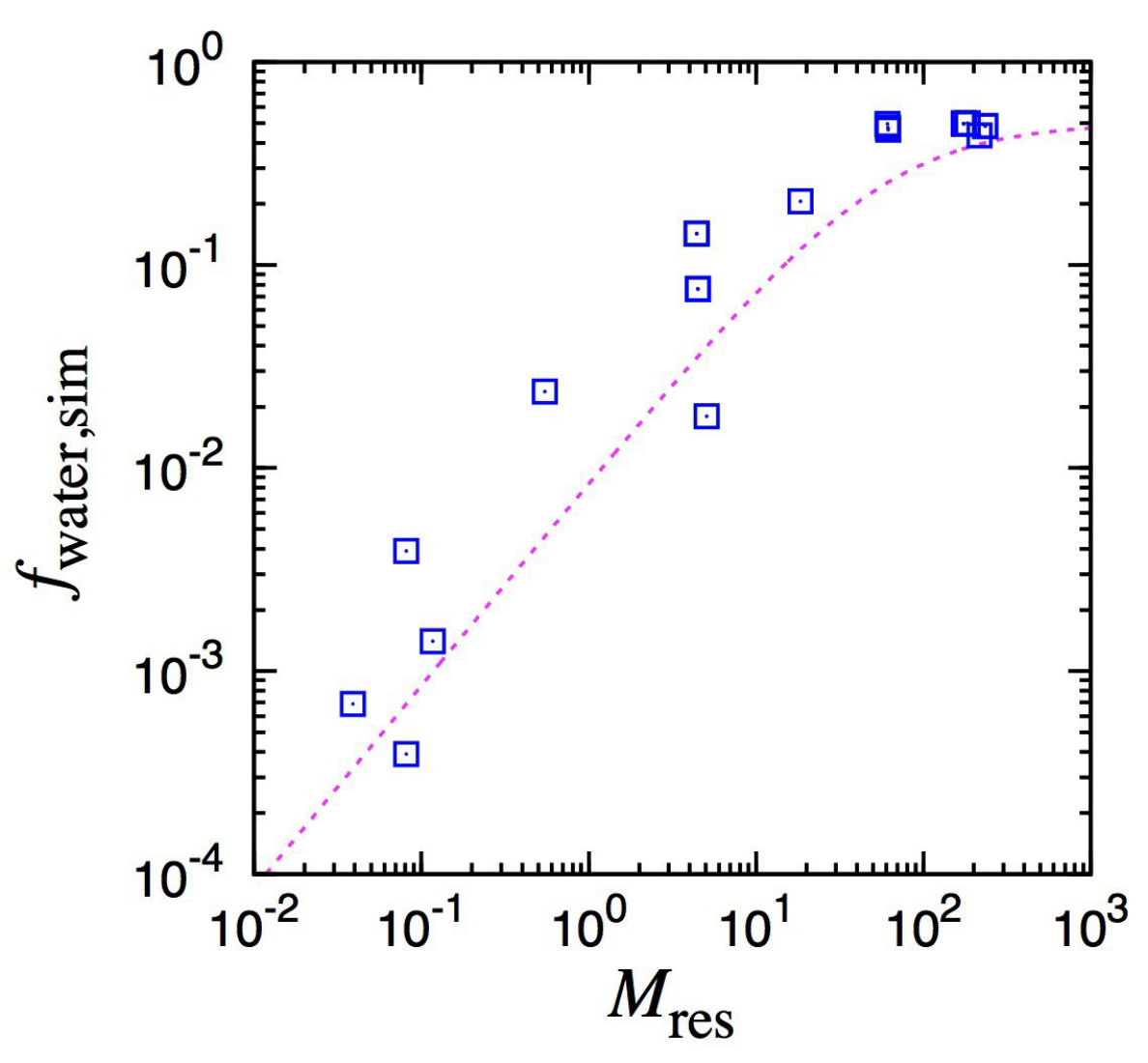}
\end{center}
\caption{The water fraction $f_{\rm water,sim}$ and 
the icy dust mass preserved in the outer disk at the snowline passage
$M_{\rm res}$ obtained by our simulations. 
The blue squares are the simulation results for the Earth analogues
with different values of $\dot{M}_{\rm g,0}, \dot{M}_{\rm pe}, t_{\rm diff}$, and $\rd0$ (section 2.4).
The dotted curve represents the analytical solution given by Eq.~(\ref{eq:fwater2}).
}\label{fig:mdust}
\end{figure*}

Equation~(\ref{eq:fwater}) is approximated to be
\begin{align}
f_{\rm water} & \simeq \frac{1}{2} \frac{f_{\rm flt}(M_{\rm pl})}{M_{\rm pl,0} + f_{\rm flt}(M_{\rm pl}) M_{\rm res}} M_{\rm res} \nonumber \\
 & \simeq 
\left\{
\begin{array}{ll}
{\displaystyle \frac{1}{2} \frac{f_{\rm flt}(M_{\rm pl,0})}{M_{\rm pl,0}} M_{\rm res}} & [\Delta M_{\rm pl} \ll M_{\rm pl,0}] \\
{\displaystyle \frac{1}{2}}  & [\Delta M_{\rm pl} \gg M_{\rm pl,0}],
\end{array}
\right.
 \label{eq:fwater1}
\end{align}
where we used $f_{\rm flt}(M_{\rm pl}) \simeq f_{\rm flt}(M_{\rm pl,0})$
for $\Delta M_{\rm pl} \ll M_{\rm pl,0}$ in the upper line.
For the range of the disk parameters we used in numerical simulations, the accretion is mostly in the 3D regime (Eq.~\ref{eq:M_2D3D})
and $f_{\rm flt} \propto M_{\rm pl}$.
In that case, Eq.~(\ref{eq:fwater1}) shows that $f_{\rm water}$
is independent of $M_{\rm pl}$ both in the cases of $\Delta M_{\rm pl} < M_{\rm pl,0}$
and $\Delta M_{\rm pl} > M_{\rm pl,0}$, while it depends on the disk parameters.
Even if the transition to 2D accretion occurs at a smaller planet mass
(which happens for smaller values of $\alpha$), 
the mass dependence of $f_{\rm water}$ is still weak: it is proportional to $M_{\rm pl}^{-1/3}$ for $\Delta M_{\rm pl} < M_{\rm pl,0}$ and independent of $M_{\rm pl}$ otherwise.
We can combine the two limits in Eq.~(\ref{eq:fwater1}) by a simple formula as 
\begin{equation}
f_{\rm water} \simeq \frac{1}{2} \frac{f_{\rm flt}(M_{\rm pl,0})}{M_{\rm pl,0} + f_{\rm flt}(M_{\rm pl,0}) M_{\rm res}} M_{\rm res}.
 \label{eq:fwater2}
\end{equation}
In Figure~\ref{fig:mdust}, $f_{\rm water,sim}$ obtained by
our simulations is compared with Eq.~(\ref{eq:fwater2})
with the simulated $M_{\rm res}$.
This figure shows that Eq.~(\ref{eq:fwater2}) reproduces the numerical results very well.
Both the numerical results and Eq.~(\ref{eq:fwater2}) show
$f_{\rm water} \propto M_{\rm res}$ for $f_{\rm water} \ll 1/2$.
The icy dust mass $M_{\rm res}$ depends sensitively on the disk parameters,
$\dot{M}_{\rm g,0}, \dot{M}_{\rm pe}, t_{\rm diff}$, and $\rd0$.
On the other hand, $f_{\rm flt}$ in 2D is independent of these disk parameters,
while that in 3D depends only weakly ($\propto \sqrt{\rd0/t_{\rm diff}}$).
In the case of $\Delta M_{\rm pl} \ll M_{\rm pl,0}$, 
$f_{\rm water} \simeq (f_{\rm flt}/M_{\rm pl,0}) M_{\rm res}$.
As we discussed, $f_{\rm flt}/M_{\rm pl,0}$ is almost constant, so that
$f_{\rm water}$ should be almost proportional to $M_{\rm res}$ for $\Delta M_{\rm pl} \ll M_{\rm pl,0}$.

\begin{figure*}\begin{center} 
\includegraphics[width=80mm]{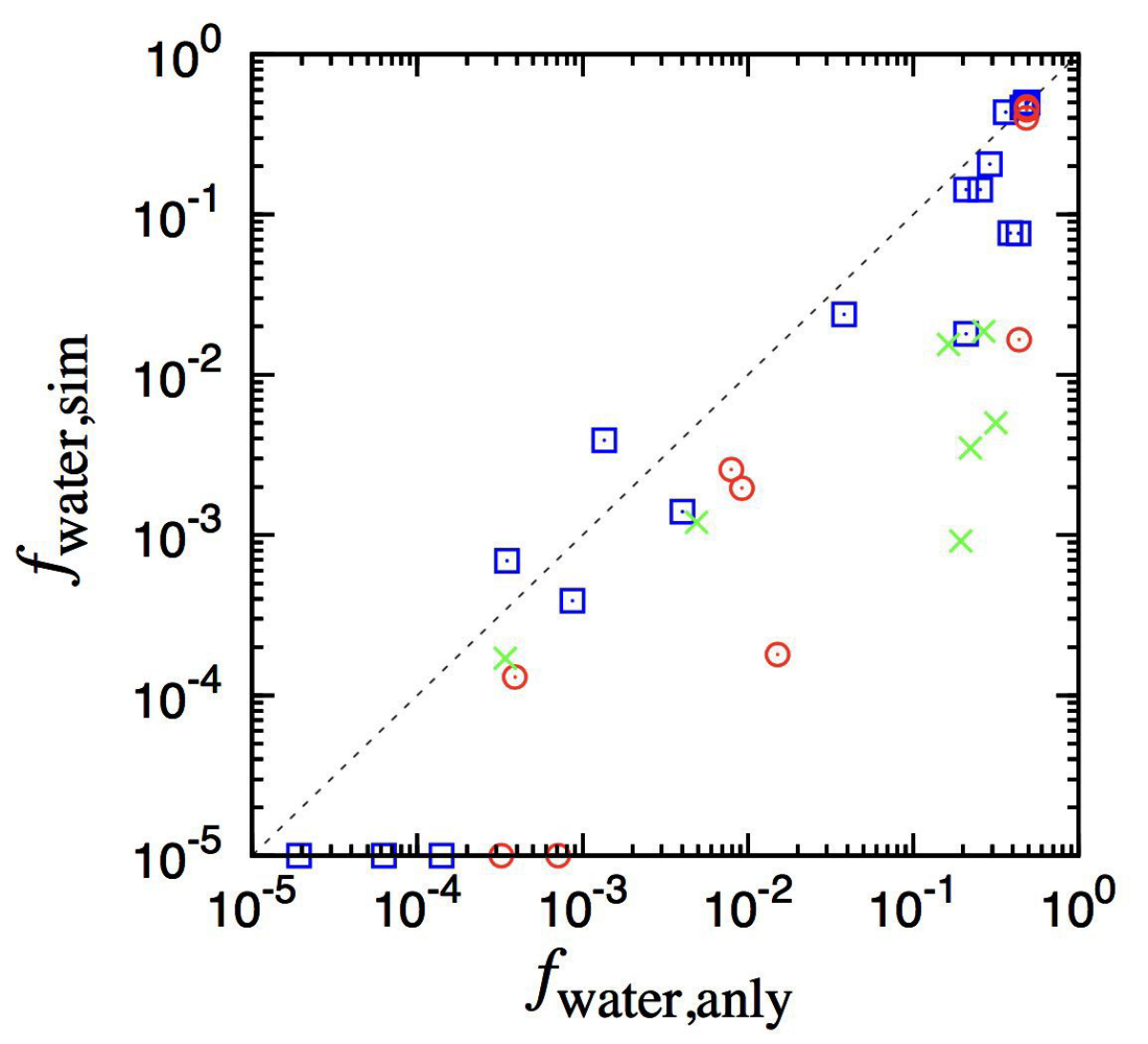}
\end{center}
\caption{Comparison of the water fraction obtained by the numerical simulations ($f_{\rm water,sim}$) with the analytical estimate ($f_{\rm water,anly}$). 
The blue squares, red circles, and green cross represent the results for the Earth, Mars, and Venus analogues, respectively. 
If $f_{\rm water,sim}$ obtained by the numerical simulation
is smaller than $10^{-5}$, we put $f_{\rm water,sim}=10^{-5}$,
because the numerical results would include a numerical uncertainty
for such small values of $f_{\rm water,sim}$.
}\label{fig:fwater_sim_anly}
\end{figure*}

Because the analytical solution given by Eq.~(\ref{eq:fwater2}) reproduces
the numerical results, 
we next derive an analytical formula for the icy dust mass $M_{\rm res}$ at the snowline passage as follows.
From Eq.~(\ref{eq:r_snow_vis}),
the snowline passes the planet orbit at $a_{\rm pl}$ when 
\begin{align}
\dot{M}_{\rm g} & \simeq \dot{M}_{\rm g, snow} \nonumber \\
  & = 2 \times 10^{-8} 
\left(\frac{a_{\rm pl}}{1\au}\right)^{9/4}\left(\frac{M_{*}}{\msol}\right)^{-3/4}
\left(\frac{\alpha}{10^{-2}}\right)^{1/2}
\msol/{\rm y}.
\end{align}
Using Eq.~(\ref{eq:alpha}),
\begin{align}
\dot{M}_{g, \rm snow} = & \, 2 \times 10^{-8} \left(\frac{M_{*}}{\msol}\right)^{-3/4} \nonumber \\ 
 & \times \left(\frac{a_{\rm pl}}{1\au}\right)^{9/4}
\left(\frac{t_{\rm diff}}{10^6 {\rm y}}\right)^{-1/2}\left(\frac{\rd0}{100 \,\au}\right)^{1/2}  
\msol/{\rm y}.
\label{eq:Mg_snow}
\end{align}
From Eq.~(\ref{eq:flux2}), the snowline passage time is
\begin{equation}
\tilde{t}_{\rm snow} \simeq \left(\frac{\dot{M}_{\rm g,0}}{\dot{M}_{g, \rm snow} + \dot{M}_{\rm pe}} \right)^{2/3};
\; \; t_{\rm snow} \simeq \left[\left(\frac{\dot{M}_{\rm g,0}}{\dot{M}_{g, \rm snow} + \dot{M}_{\rm pe}} \right)^{2/3} - 1 \right] t_{\rm diff}.
\label{eq:tsnow}
\end{equation}
From Eq.~(\ref{eq:M_g2}), the remaining gas disk mass at the snowline passage is given by
\begin{equation}
M_{\rm g,snow} \simeq 2 \, \tilde{t}_{\rm snow} \, t_{\rm diff}
\left( \dot{M}_{\rm g,0} \, \tilde{t}_{\rm snow}^{\;\, -3/2} - \dot{M}_{\rm pe} \right).
\end{equation}
Because $M_{\rm res} \sim Z \,M_{\rm g,snow}$, we obtain 
\begin{equation}
M_{\rm res} \sim 2 \, Z_0 \left(1 + \frac{t_{\rm snow}}{t_{\rm pff}}\right)^{-\gamma} 
\, \tilde{t}_{\rm snow} \, t_{\rm diff}
\left( \dot{M}_{\rm g,0} \, \tilde{t}_{\rm snow}^{\;\, -3/2} - \dot{M}_{\rm pe} \right),
\label{eq:mdust}
\end{equation}
where we approximated $Z$ as $Z \sim (1+t/t_{\rm pff})^{-\gamma} \, Z_0$ from
Eq.~(\ref{eq:Sigma_decay}), $t_{\rm pff} \sim 2 \times 10^5 (r/100\,{\rm au})^{3/2} (M_*/M_\odot)^{-1/2}\, {\rm y}$
(Eq.~\ref{eq:tpff}), and
$\gamma = 1 + 0.15 (300\,{\rm au}/\rd0)$ (Eq.~\ref{eq:gamma2}).
Substituting the filtering factor $f_{\rm flt}$ given by Eq.~(\ref{eq:fflt}) with St = 0.1 and $M_{\rm res}$ estimated above into Eq.~(\ref{eq:fwater2}), we can analytically estimate the water fraction from the disk parameters, $\dot{M}_{\rm g,0}, \dot{M}_{\rm pe}, t_{\rm diff}$ and $r_{\rm d,0}$ as (Eqs.~\ref{eq:fflt} and ~\ref{eq:fwater})
\begin{align}
f_{\rm water} \simeq \frac{1}{2}
\left(1+\frac{M_{\rm pl,0}}{f_{\rm flt}(M_{\rm pl,0}) M_{\rm res}} \right)^{-1},
\label{eq:fwater3}
\end{align}
where $M_{\rm res}$ is given by Eq.~(\ref{eq:mdust}),
$f_{\rm flt} = \min(f_{\rm flt,3D}, f_{\rm flt,2D})$, and
\begin{align}
f_{\rm flt,3D} \simeq & \;
0.017\left( \frac{M_{\ast}}{M_{\odot}} \right)^{-1} \left(\frac{\alpha}{10^{-2}} \right)^{-1/2} 
\left(\frac{h_{\rm g}/r}{0.02}\right)^{-1} 
\left(\frac{\rm St}{0.1}\right)^{1/2} \nonumber\\
 & \times
\left(\frac{M_{\rm pl,0}}{0.1 M_\oplus}\right)
\left(\frac{r}{1 {\rm au}} \right)^{-1/2},  \\
f_{\rm flt,2D} \simeq & \;
0.040 \left( \frac{M_{\ast}}{M_{\odot}} \right)^{-2/3} 
\left(\frac{\rm St}{0.1}\right)^{-1/3} 
\left(\frac{M_{\rm pl,0}}{0.1 M_\oplus}\right)^{2/3} 
\left(\frac{r}{1 {\rm au}} \right)^{-1/2}.  
\end{align}
Except for the value of $\gamma$, which was fitted by the numerical results,
the other derivations are analytical.
Note that $f_{\rm water} \propto M_{\rm res} \propto Z_0$.
Around metal-rich stars, $f_{\rm water}$ would be larger.

In Figure \ref{fig:fwater_sim_anly},
we compare the analytically estimated water fraction $f_{\rm water,anly}$
with the numerically simulated $f_{\rm water,sim}$.
In addition to the Earth analogues ($M_{\rm pl,0} = 1 M_\oplus$) at 1 au,
we also plot the results 
for the Mars analogues ($M_{\rm pl,0} = 0.11 M_\oplus$) at 1.52 au
and the Venus analogues ($M_{\rm pl,0}= 0.82 M_\oplus$) at 0.72 au.
For the Earth and Mars analogues,
$f_{\rm water,anly}$ reproduces $f_{\rm water,sim}$
within a factor of several, while the water fraction varies
by several orders of magnitude, except for some runs where the water fraction 
is overestimated by the analytical formula.
The agreement both for the Earth and Mars analogues 
strongly suggests that the mass and semimajor axis dependences
are also reproduced.
In the case of Venus analogues, 
$f_{\rm water,anly}$ is larger by a factor of a few to a few tens than $f_{\rm water,sim}$.
When the Earth analogue increases its mass by capture of pebbles, 
it captures significant fraction of the pebble flux and 
the accretion of the Venus analogues that resides in inner region of the Earth analogues
can be significantly decreased.
This effect is included in the numerical simulation, while
it is not taken into account in the analytical formula.

\section{Dependence of water fraction on disk parameters}

\begin{figure*}\begin{center} 
\includegraphics[width=150mm]{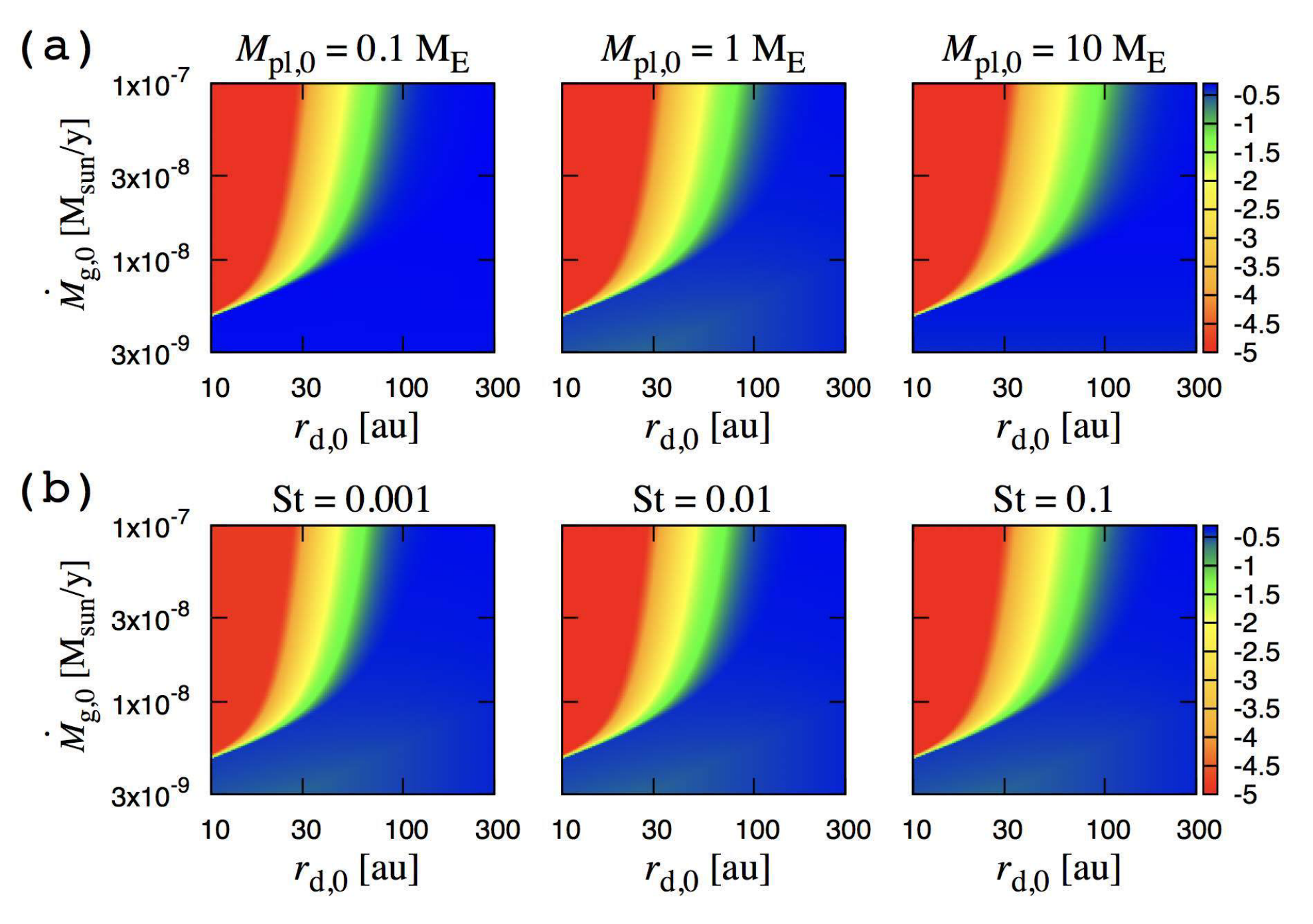}
\end{center}
\caption{The analytical estimate of $f_{\rm water}$ at 1 au
as a function of the initial disk radius ($r_{\rm d,0}$) and 
the initial disk accretion rate ($\dot{M}_{\rm g,0}$).
(a) The dependence on the planet mass ($M_{\rm pl,0}$),
and (b) that on the Stokes number (St) of pebbles.
The other parameters are $\dot{M}_{\rm pe} = 10^{-9} M_{\odot}/{\rm y}$
and $t_{\rm diff} = 3 \times 10^6{\rm y}$.
In the panels (a), we used St = 0.1.
In the panels (b), $M_{\rm pl,0}=1 M_\oplus$.
The middle panels in (a) and (b) are identical. 
The color scales are $\log_{10} f_{\rm water}$.
}\label{fig:water_M_St}
\end{figure*}

Using the analytical formula, we investigate
how the water fraction ($f_{\rm water}$) is
determined by the disk parameters.
Figure~\ref{fig:water_M_St} shows $f_{\rm water}$ for a planet at $a_{\rm pl}=1.0$ au as a function of
the disk parameters, $\dot{M}_{\rm g,0}$ and $r_{\rm d,0}$.
The other parameters are $\dot{M}_{\rm pe} = 10^{-9} M_{\odot}/{\rm y}$
and $t_{\rm diff} = 3 \times 10^6{\rm y}$.
The panels (a) show the dependence on $M_{\rm pl,0}$ for St = 0.1.
The panels (b) show the dependence on St for $M_{\rm pl,0}=1 M_\oplus$.
The planets formed with the parameters in the red region are very dry, $f_{\rm water} \la 10^{-4}$.
Those in the green region have modest amount of water, $f_{\rm water} \sim 0.1$,
and those in the blue region are icy planets, $f_{\rm water} \simeq 1/2$.
The yellow and orange regions represent $f_{\rm water} \sim 10^{-4} - 10^{-2}$,
which corresponds to the water fraction of the current Earth
and that estimated for the ancient Mars.

\begin{figure*}\begin{center} 
\includegraphics[width=150mm]{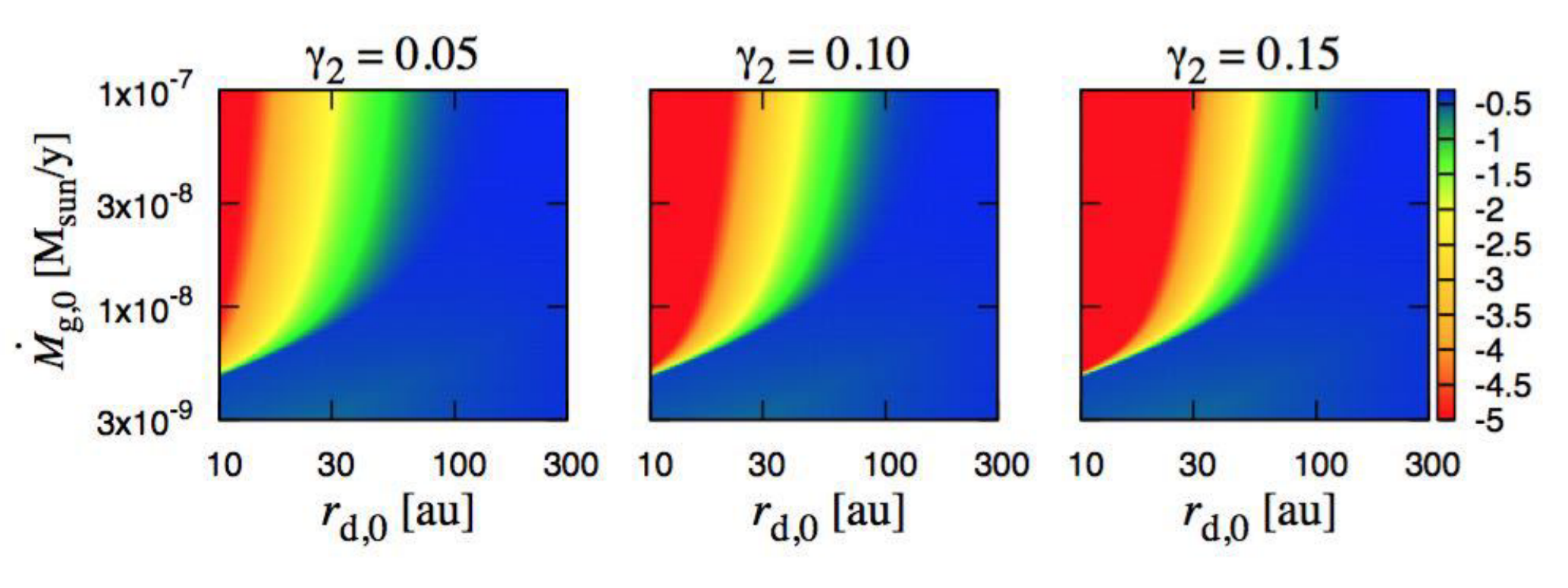}
\end{center}
\caption{The analytical estimate of $f_{\rm water}$ at 1 au
as a function of the initial disk radius ($r_{\rm d,0}$) and 
the initial disk accretion rate ($\dot{M}_{\rm g,0}$),
with $\gamma_2 = 0.05, 0.10$, and 0.15.
The other parameters are ${\rm St} = 0.1$,
$M_{\rm pl,0}=1 M_\oplus$, $\dot{M}_{\rm pe} = 10^{-9} M_{\odot}/{\rm y}$,
and $t_{\rm diff} = 3 \times 10^6{\rm y}$.
The color scales are $\log_{10} f_{\rm water}$.
}\label{fig:water_gamma}
\end{figure*}

\begin{figure*}\begin{center} 
\includegraphics[width=150mm]{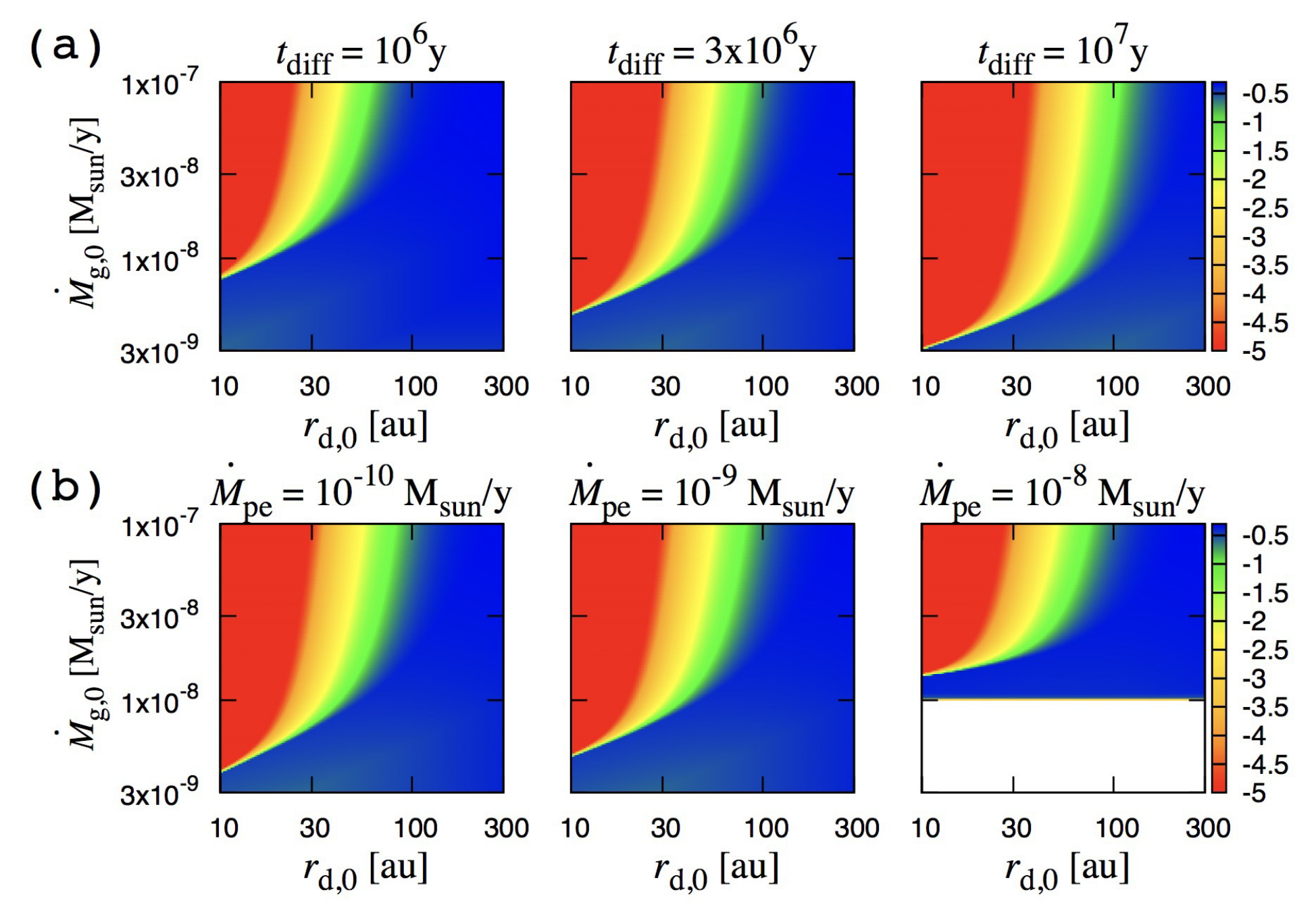}
\end{center}
\caption{The analytical estimate of $f_{\rm water}$ at 1 au
as a function of $r_{\rm d,0}$ and $\dot{M}_{\rm g,0}$.
(a) The dependence on the disk diffusion timescale $t_{\rm diff}$,
and (b) that of the photoevaporation rate ($\dot{M}_{\rm pe}$).
The other parameters are ${\rm St} = 0.1$
and $M_{\rm pl,0}=1 M_\oplus$.
In the panels (a), we used $\dot{M}_{\rm pe} = 10^{-9} M_{\odot}/{\rm y}$,
and $t_{\rm diff} = 3 \times 10^6{\rm y}$ in the panels (b).
The middle panels of (a) and the right panel of (b) are identical. 
In the right panel of (b), the region with $\dot{M}_{\rm g,0} < \dot{M}_{\rm pe}$
is empty, because the disks do not exist under that condition. 
The color scales are $\log_{10} f_{\rm water}$.
}\label{fig:water_tdiff_pe}
\end{figure*}

We find that $f_{\rm water}$ is the most sensitive to $\dot{M}_{\rm g,0}$ and $r_{\rm d,0}$.
The common features in the contours in Fig.~\ref{fig:water_M_St} are
that i) $f_{\rm water}$ is lower for smaller $r_{\rm d,0}$ and larger $\dot{M}_{\rm g,0}$, 
and ii) $f_{\rm water} \sim 10^{-4} - 10^{-2}$
is realized at $r_{\rm d,0} \sim 30-50\, {\rm au}$ and 
$\dot{M}_{\rm g,0} \ga 10^{-8} M_{\odot}/{\rm y}$.
This is consistent with the conclusion by \cite{Sato2016} that
such $f_{\rm water}$ can be realized for compact disks with sizes $< 100 \, {\rm au}$
and late snowline passage ($> 2-4$ My).
We here show that the disk parameters for 
$f_{\rm water} \sim 10^{-4} - 10^{-2}$ is not in a very narrow window.

As we showed in section 4, 
$f_{\rm water}$ is regulated by $M_{\rm res}$ and
$M_{\rm res}$ is sensitive to $\dot{M}_{\rm g,0}$ and $r_{\rm d,0}$, especially through the parameter 
$t_{\rm snow}/t_{\rm pff}$ (Eq.~\ref{eq:mdust}).
Because pebble accretion is fast, in order to realize $f_{\rm water} \sim 10^{-4} - 10^{-2}$,
$M_{\rm res}$ has to have significantly decayed until $t \sim t_{\rm snow}$.
For small $r_{\rm d,0}$, $t_{\rm pff}$ is small (Eq.~{\ref{eq:tpff}),
while $t_{\rm snow}$ is large (Eq.~{\ref{eq:tsnow}) due to
small $\dot{M}_{\rm g,snow}$ (Eq.~{\ref{eq:Mg_snow}).
For large $\dot{M}_{\rm g,0}$, $t_{\rm snow}$ is small (Eq.~{\ref{eq:Mg_snow})
while $t_{\rm pff}$ is the same (Eq.~{\ref{eq:tpff}).
As a result, $\Sigma_{\rm p}$ decays more quickly (Eq.~\ref{eq:Sigma_decay})
for small $r_{\rm d,0}$ and large $\dot{M}_{\rm g,0}$.

Figure~\ref{fig:water_M_St} shows the dependences of
$f_{\rm water}$ on 
the pebble accretion parameters: the initial planetary mass ($M_{\rm pl,0}$)
and the Stokes number (St) of pebbles that accrete onto the planet.
In the numerical simulations, the Stokes number of pebbles that
accrete onto the planet is calculated
by growth and radial drift of pebbles in an evolving disk.
According to the simulations, the Stokes number of radially drifting particles is $\sim 0.1$ at early times (Eq.~\ref{eq:eqSt}), but decreases with time as the dust and gas disks evolve. 
As Eq.~(\ref{eq:Sigma_decay}) shows, $\Sigma_{\rm p}$ decays more rapidly
than $\Sigma_{\rm g}$ for $t > t_{\rm pff}$.
Accordingly, $Z$ decreases and 
the equilibrium Stokes number of migrating pebbles become smaller 
(Eq.~\ref{eq:eqSt}; also see \citet{Sato2016}).
As we have pointed out in section 3.1, if a fragmentation/rebound barrier
limits the icy pebble growth, St also becomes small.
By these reasons, we also showed plots with St = 0.01 and 0.001
in Fig.~\ref{fig:water_M_St}.

Figure~\ref{fig:water_M_St} shows that
$f_{\rm water}$ is almost independent of $M_{\rm pl,0}$ and St.
The weak dependence on $M_{\rm pl,0}$
is explained by the following arguments.
The water fraction $f_{\rm water}$ is 
$\sim (f_{\rm flt}/M_{\rm pl,0}) M_{\rm res}$ in the case of $\Delta M_{\rm pl} \ll M_{\rm pl,0}$.
The dust mass $M_{\rm res}$ is independent of $M_{\rm pl}$.
The factor $f_{\rm flt}/M_{\rm pl,0}$ is independent of $M_{\rm pl,0}$ in 3D accretion
and weakly depends on $M_{\rm pl,0}$ in 2D accretion ($\propto M_{\rm pl,0}^{-1/3}$),

The weak dependence of $f_{\rm water}$ on Stokes number 
shown in Figure~\ref{fig:water_M_St} is resulted by the assumption that $M_{\rm res}$ is independent of St;
it depends on St only weakly (Eq.~\ref{eq:fflt}) through $f_{\rm flt}$.
The parameter $\gamma$ could be dependent on St, because
the sculpture rate of $\Sigma_{\rm p}$ at $r=r_{\rm pff}$ may depends on St.
Figure~\ref{fig:water_gamma} shows the dependence on $\gamma_2$,
assuming that the functional form of 
$\gamma = 1+ \gamma_2(300{\rm au}/\rd0)$ still holds. 
Because the sculpture rate may be lower for smaller values of St,
we tested the cases of $\gamma_2 = 0.05$ and 0.1, in addition to
the nominal case of $\gamma_2 = 0.15$.
The result is insensitive to $\gamma_2$ for $\gamma_2 \ga 0.1$,
while the sculpture rate is lower and accordingly the $f_{\rm water}$ is relatively higher
for $\gamma_2 \sim 0.05$.
The detailed functional form of $\gamma$ is left for future work.

\begin{figure*}\begin{center} 
\includegraphics[width=150mm]{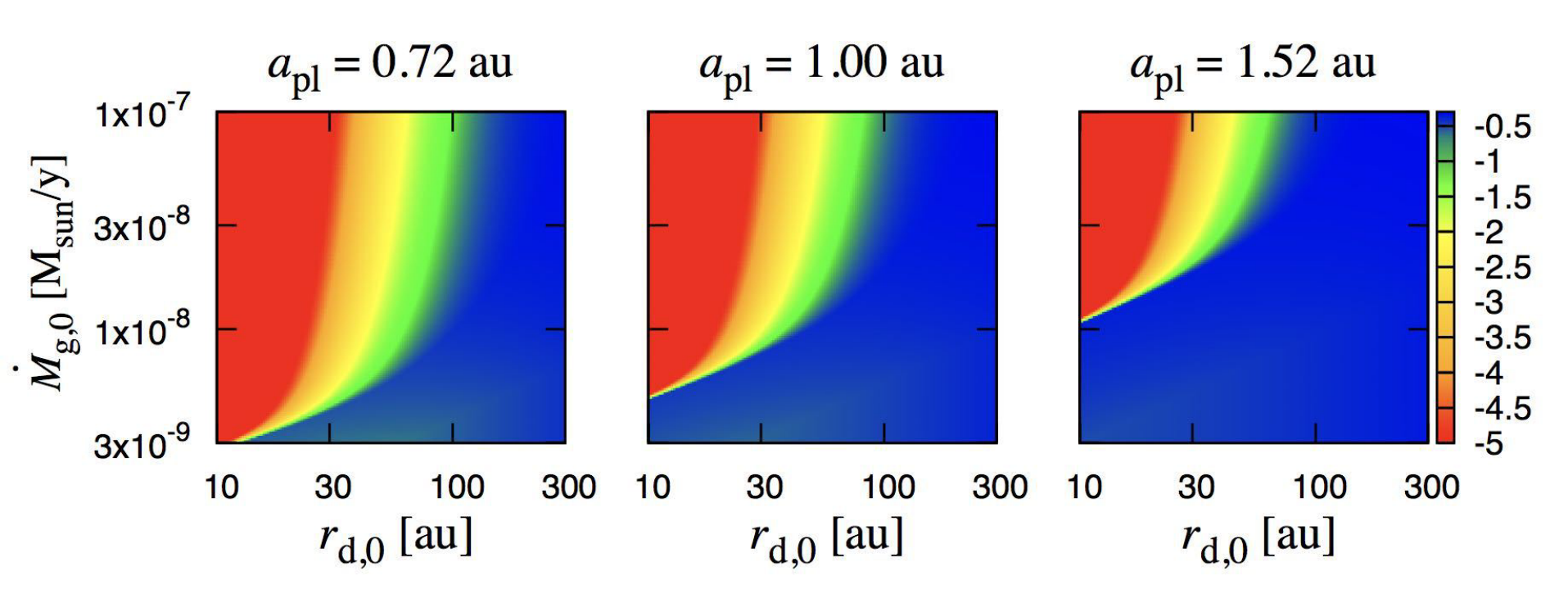}
\end{center}
\caption{The analytical estimate of $f_{\rm water}$ 
as a function of $r_{\rm d,0}$ and 
$\dot{M}_{\rm g,0}$, at 0.72, 1.00, and 1.52 au.
The other parameters are ${\rm St} = 0.1$,
$M_{\rm pl,0}=1 M_\oplus$, $\dot{M}_{\rm pe} = 10^{-9} M_{\odot}/{\rm y}$,
and $t_{\rm diff} = 3 \times 10^6{\rm y}$.
The color scales are $\log_{10} f_{\rm water}$.
}\label{fig:water_ap}
\end{figure*}

\begin{figure*}\begin{center} 
\includegraphics[width=150mm]{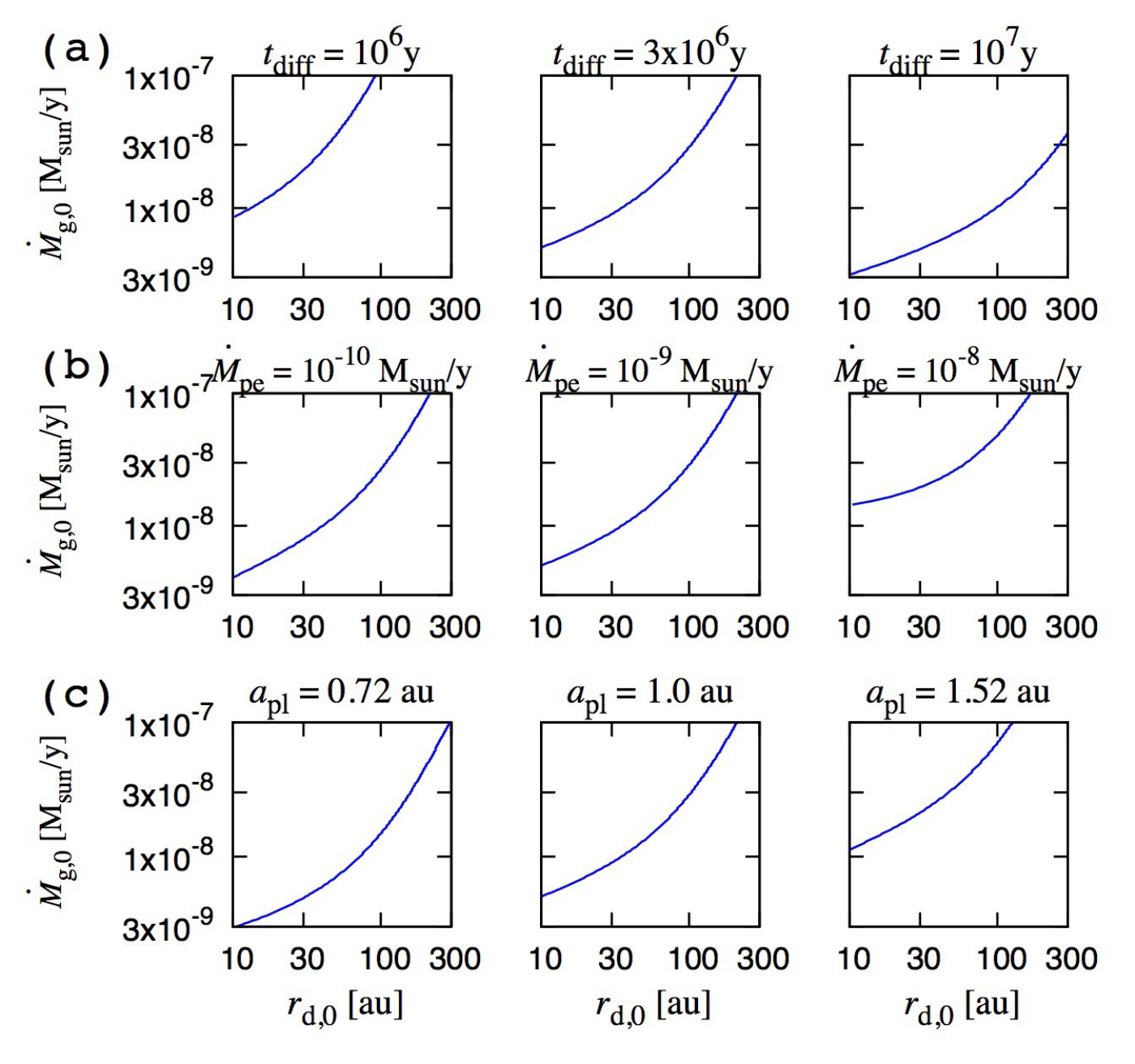}
\end{center}
\caption{
The condition of
$t_{\rm snow} = 10\, t_{\rm pff}$ as a function of $r_{\rm d,0}$ and $\dot{M}_{\rm g,0}$.
(a) The dependence on the disk diffusion timescale $t_{\rm diff}$,
(b) that of the photoevaporation rate ($\dot{M}_{\rm pe}$),
and (c) that of planetary orbital radius ($a_{\rm pl})$.
The right regions from the curves represent water-rich regions
($f_{\rm water}\sim 1/2$).
In the panels (a), we used $\dot{M}_{\rm pe} = 10^{-9} M_{\odot}/{\rm y}$
and $a_{\rm pl}=1$ au,
$t_{\rm diff} = 3 \times 10^6{\rm y}$ and $a_{\rm pl}=1$ au in the panels (b),
and $\dot{M}_{\rm pe} = 10^{-9} M_{\odot}/{\rm y}$ and $t_{\rm diff} = 3 \times 10^6{\rm y}$ in the palens (c).
}\label{fig:crt}
\end{figure*}

Figure~\ref{fig:water_tdiff_pe} shows 
the dependence on the other disk parameters, 
$t_{\rm diff}$ and $\dot{M}_{\rm pe}$.
For larger $t_{\rm diff}$, the 
evolution of the snowline is slower and the snowline passage is later (Eq.~\ref{eq:tsnow}).
Before the passage, the icy dust disk has been more sculpted.
As a result, $M_{\rm res}$ is smaller (Eq.~\ref{eq:mdust}), 
because $t_{\rm snow} \propto t_{\rm diff}$ and $\gamma \ge 1$.
For smaller $\dot{M}_{\rm pe}$, 
the disk is hotter and the snowline passage is later ($t_{\rm snow}$ is larger),
resulting in smaller $M_{\rm res}$.
If $M_{\rm res}$ is smaller, $f_{\rm water}$ is smaller.
The water fraction corresponding to the current Earth is $10^{-4} - 10^{-2}$
(the yellow and orange colored regions) are only slightly shifted
to larger $r_{\rm d,0}$ and lower $\dot{M}_{\rm g,0}$
for smaller $t_{\rm diff}$ and larger $\dot{M}_{\rm pe}$,
because $M_{\rm res}$ is smaller for these parameters (Eq.~\ref{eq:mdust}).
Note that the parameter region with $\dot{M}_{\rm g,0} < \dot{M}_{\rm pe}$
is empty in the right panel of Figure~\ref{fig:water_tdiff_pe}b, because the disks do not exist under that condition.

In Figure \ref{fig:water_ap}, the analytically estimated $f_{\rm water}$ is plotted
for $a_{\rm pl} = 0.72, 1.00$ and 1.52 au (the Venus, Earth, and Mars analogues, respectively).
In the outer region, $f_{\rm water}$ is generally larger due to
early snowline passage (smaller $t_{\rm snow}$).
In other words,  $\dot{M}_{\rm g,snow}$ is larger for larger $a_{\rm pl}$.
As mentioned in section 4, the Venus may have further lower $f_{\rm water}$,
if we take into account the decrease in the pebble flux due to accretion by the Earth analogue.
As shown in Eq.~(\ref{eq:r_snow_irr}), the snowline cannot reach the region inside 0.53 au
in our disk model, the planets there are completely dry.
Although the dependence on the orbital radius exists, 
in the case of $r_{\rm d,0} \sim 30-50\, {\rm au}$ and $\dot{M}_{\rm g,0} \ga 2 \times 10^{-8} M_{\odot}/{\rm y}$,
$f_{\rm water} \sim 10^{-4} - 10^{-2}$ both for the Earth and Mars analogues.

In section 3, we pointed out that the simple condition 
of $t_{\rm snow}/t_{\rm pff} < 10$ or $>10$ discriminates
between the water-rich case ($f_{\rm water}\sim 1/2$) and the water-poor case.
The pebble formation timescale $t_{\rm pff}$ is given by Eq.~(\ref{eq:tpff})
and the snowline passage time $t_{\rm snow}$ is given by Eq.~(\ref{eq:tsnow}).
Both $t_{\rm pff}$ and $t_{\rm snow}$ are independent of
$M_{\rm pl}$ and St, which is consistent with Figs.~\ref{fig:water_M_St}.
The condition of $t_{\rm snow} = 10 \, t_{\rm pff}$ is
shown on the $\rd0$-$\dot{M}_{\rm g,0}$ plane in Fig.~\ref{fig:crt}.
We show the dependences on $t_{\rm diff}$, $\dot{M}_{\rm pe}$, 
and $a_{\rm pl}$.
Comparison of this figure with Figures~\ref{fig:water_tdiff_pe} and \ref{fig:water_ap}
show that the simple condition approximately reproduces the more detailed
evaluation except for $\rd0 > 100$ au.
As shown in Fig.~\ref{fig:lines}, for $\rd0 > 100$ au,
the disk radius expands with the pebble formation front radius $r_{\rm pff}$,
so that the sculpture of icy dust reservoir is delayed, compared to 
the estimation of $t_{\rm pff}$ at $\rd0$.
The calculation of $M_{\rm res}$ does not have this problem.

As we have shown,
$f_{\rm water}$ is the most sensitive to $\dot{M}_{\rm g,0}$ and $r_{\rm d,0}$
and almost independent of the other parameters of disks and pebble accretion.
It would be a robust result that the water fraction inferred for 
the present Earth and the ancient Mars, $f_{\rm water} \sim 10^{-4} - 10^{-2}$, 
is realized at $r_{\rm d,0} \sim 30-50\, {\rm au}$ and 
$\dot{M}_{\rm g,0} \ga 2 \times 10^{-8} M_{\odot}/{\rm y}$,
which may correspond to median disk of classical T Tauri stars or
slightly compact and massive disks.

\section{Discussion}

\begin{figure*}\begin{center} 
\includegraphics[width=150mm]{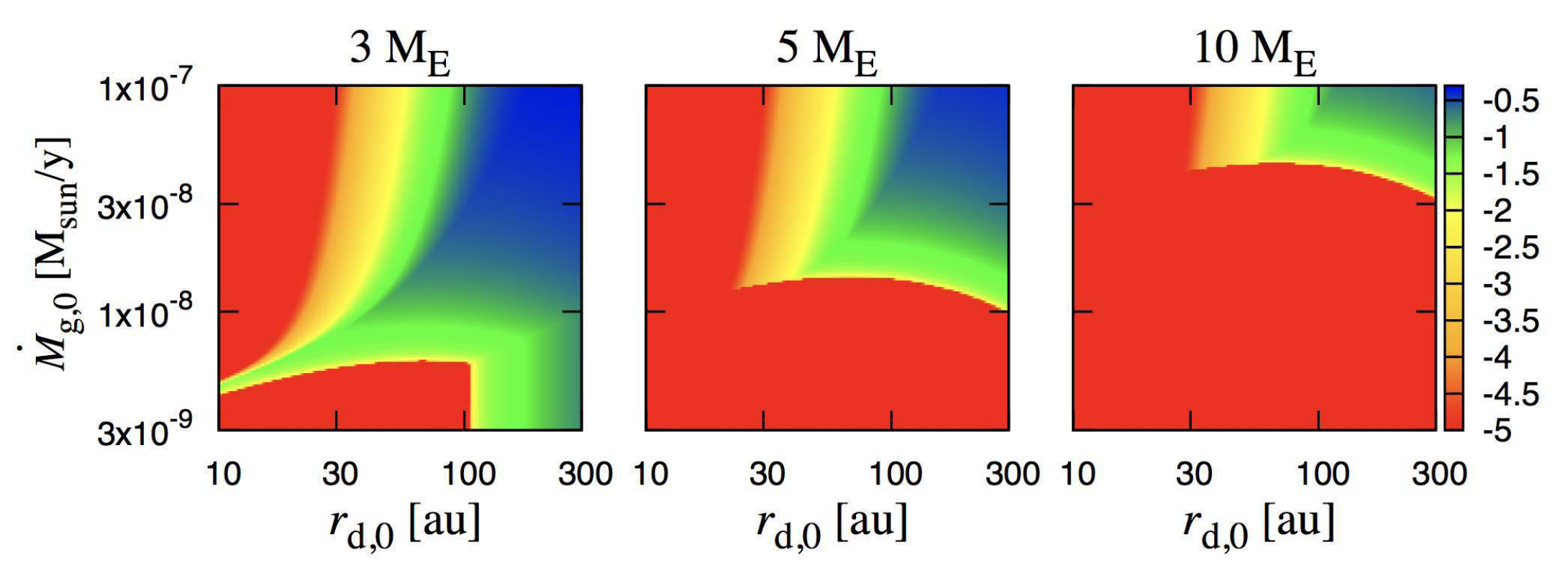}
\end{center}
\caption{The analytical estimate of $f_{\rm water}$ at 1 au
as a function of the initial disk radius ($r_{\rm d,0}$) and 
the initial disk accretion rate ($\dot{M}_{\rm g,0}$),
including the effect of the pebble isolation mass. 
The other parameters are $\dot{M}_{\rm pe} = 10^{-9} M_{\odot}/{\rm y}$,
$t_{\rm diff} = 3 \times 10^6{\rm y}$, and St = 0.1.
The pebble accretion is halted when $M_{\rm pl}$ reaches $M_{\rm peb,iso}$.
The dry regime (the red-colored regime) in the left bottom part of 
the plots represent the cases of $M_{\rm pl,0} > M_{\rm peb,iso}$.
}\label{fig:water_Miso}
\end{figure*}

Here we comment on the effect of pebble isolation mass,
which we did not include in our simulations.
A planet with relatively large mass ($M_{\rm pl}$)
can make a dip in the gas disk along the planetary orbit
to prevent the pebbles from passing the orbit.
The threshold mass is called ''pebble isolation mass" ($M_{\rm peb,iso}$)
\citep{Lambrechts2014,Bitsch2018,Ataiee18}.
When the planetary mass reaches the isolation mass,
the pebble accretion onto the planet with $M_{\rm pl} > M_{\rm peb,iso}$
and other planets inside the planetary orbit is truncated  
and the increases in their water fraction are stalled.

\citet{Bitsch2018} derived a detailed expression of the
pebble isolation mass,
\begin{equation}
M_{\rm peb,iso} \simeq 25 \left(\frac{h/r}{0.05}\right)^3
\left[0.34 \left(\frac{3}{\log_{10}(\alpha)}\right)^4 + 0.66\right] M_{\oplus}.
\end{equation}
For example,
$M_{\rm peb,iso} \simeq 4.1 M_\oplus$ for 
$t_{\rm diff} = 3\times 10^6{\rm y}, \dot{M}_{\rm g,0}=10^{-8}M_\odot/{\rm y}$
and $\rd0=100$ au,
$h/r \simeq 0.0246$ and $\alpha \simeq 3 \times 10^{-3}$ at 1 au.
If $M_{\rm pl,0} \ga {\rm several}\; M_\oplus$,
the effect of pebble isolation mass is not negligible. 
Figure~\ref{fig:water_Miso} shows
the water fraction for $M_{\rm pl,0} = 3, 5$, and $10 M_\oplus$.
We stop the pebble accretion when $M_{\rm pl}$ reaches $M_{\rm peb,iso}$.
Because $\alpha$ is smaller for smaller $\rd0$ (Eq.~\ref{eq:alpha})
and $h/r$ is lower for smaller $\dot{M}_{\rm g,0}$ and higher $\alpha$
(Eq.~\ref{eq:hvis}), $M_{\rm peb,iso}$ is smaller for 
smaller $\rd0$ and $\dot{M}_{\rm g,0}$.
The dry regime (the red-colored regime) in the left bottom part of 
the plots represent the cases of $M_{\rm pl,0} > M_{\rm peb,iso}$.
Because $t_{\rm snow}$ is small in the low $\dot{M}_{\rm g,0}$ regions,
$f_{\rm water}$ rapidly increases due to a high pebble flux until $M_{\rm pl}$ increases to $M_{\rm peb,iso}$.
Thereby, the edge in $f_{\rm water}$ at $M_{\rm pl,0} \simeq M_{\rm peb,iso}$ is sharp.
ion is avoided by the effect of pebble isolation even after the snowline passage.

For the same $t_{\rm diff}$, $\dot{M}_{\rm g,0}$, and $\rd0$ as the above,
$M_{\rm peb,iso} \simeq 4.1 (r/1\,{\rm au})^{6/7} M_\oplus$.
The filtering rate for $M_{\rm peb,iso}$ is $f_{\rm flt} \simeq 0.5 (r/1\,{\rm au})^{1/14}$
 in 2D case (Eq.~\ref{eq:fflt}).
 If ice giants or cores of gas giants are formed,
 even before their mass exceeds $M_{\rm peb,iso}$, the pebble flux
 is reduced by $\sim 50\%$ by individual ice giants.

If Jupiter's core is formed in the outer region,
it shuts down the pebble mass flux into the terrestrial planet region.
\cite{Morbidelli2016} proposed that the pebble flux truncation by 
the formation of Jupiter accounts for the dichotomy of our Solar system -- 
the total solid mass contained in Jupiter, Saturn, Uranus and Neptune is
about 50 times larger than the total mass of terrestrial planets.
Jupiter's core must be formed
at $M_{\rm res} \ga 10 M_\oplus$.
From Eq.~(\ref{eq:mdust}) with $M_{\rm d,0} \sim 10^{-8}M_\odot/{\rm y}$
and $t_{\rm diff}\sim 3\times 10^6$ y, 
$M_{\rm res}(t) \sim 200 (t_{\rm pff}/t)^\gamma M_\oplus$,
which is $M_{\rm res}(t) \sim 20 (t/10^6\,{\rm y})^{-1.5} M_\oplus$
for $\rd0 \sim 100$ au.
Hence, the Jupiter's core formation time $t_{\rm jup}$ must be $\la t_{\rm diff}$.
While too fast type I migration of the core is a problem in that case \citep{Matsumura2017}, the dichotomy of our Solar system could be created.
Our results show that $f_{\rm water}$ rapidly increases
jafter $t = t_{\rm snow}$ and becomes saturated before $t = t_{\rm diff}$.
If $t_{\rm snow}> t_{\rm jup}$, $f_{\rm water}=0$ for terrestrial planets.
Otherwise, it is likely that $f_{\rm water}$ is already close to the
saturated value, because the increase of $f_{\rm water}$ is very rapid.
It is difficult for giant planet formation to directly 
produce modestly low values of water 
mass fraction ($\sim 10^{-4}-10^{-2}$) corresponding to the Earth and ancient Mars.
The modestly low values are attained by 
disk parameters with $t_{\rm snow} \sim 10 \, t_{\rm pff}$ as we showed.

We also point out that D/H ratio is expected to be radially uniform among the Earth,
asteroids, and comets, 
if water is delivered by icy pebbles that formed in disk outer regions
and drift all the way to the host star.
However, observations show that in our Solar system, D/H ratios 
of Oort cloud comets are clearly higher than the Earth \citep[e.g.,][]{Marty2012}.
One possibility to reconcile the discrepancy is the shut-down of the pebble flux
by Jupiter formation \citep{Kruijer2017}.
Because pebble formation front migrates outward, 
the isotope ratios of drifting pebbles before and after Jupiter formation,
which are respectively 
the building materials for terrestrial planets and those for comets,
should be different.
More careful comparison should be necessary between planet formation model and
cosmo-chemical data.

\section{Summary}

If water is not delivered to rocky planets in habitable zones,
the planets cannot be actual habitats, because   
H$_2$O ice condenses in the disk regions well beyond the habitable zones.
In the pebble accretion model, accretion of
icy pebbles after the snowline passage 
may be a primary mechanism to deliver water to the rocky planets. 

In this paper, we have investigated the water delivery to rocky planets
by pebble accretion around solar-type stars, through 1D simulation of
the growth of icy dust grains to pebbles and the pebble 
radial drift in an evolving disk.
We assume that the planetary embryos did not migrate significantly
and consist of pure rock,
which means that accretion of ice starts when the snowline migrates
inward and passes the planetary orbit due to disk evolution.
Our previous paper, \citet{Sato2016}, pointed out that the water fraction of the final planets
are determined by the timings of the snowline passage through the planetary orbit ($t = t_{\rm snow}$) and disk gas depletion, because pebble accretion is fast and efficient.
While \citet{Sato2016} used a simple static disk model,
we here used the evolving disk model due to viscous diffusion 
based on the self-similar solution with constant viscous $\alpha$
\citep{LyndenPringle1974}
and simultaneously calculated pebble formation/drift/accretion
and snowline migration with the disk model.
Because the snowline migration is correlated with global disk diffusion
in the evolving disk model,
we found that for water fraction of the final planets ($f_{\rm water}$),
the snowline passage time ($t_{\rm snow}$)
relative to the time ($t_{\rm pff}$) at which pebble formation front reaches the disk outer edge
is more important than that relative to the disk gas depletion timescale.
Our simulation shows that the ice dust mass preserved in the disk outer region 
at $t = t_{\rm snow}$ ($M_{\rm res}$) determines the water fraction of the final planets.
The accreted ice mass to the planet is estimated by $\sim (1/2) f_{\rm flt} M_{\rm res}$, where the filtering factor $f_{\rm flt}$ is the fraction of the pebble mass flux that
is accreted onto the planet. 
Because $M_{\rm res}$ rapidly decreases after $t \sim t_{\rm pff}$, 
$t_{\rm snow}/t_{\rm pff} > 10$ or $< 10$ is crucial for final value of $f_{\rm water}$.
If $t_{\rm snow}/t_{\rm pff} > 10$, $M_{\rm res}$ should have 
significantly decayed 
when icy pebble accretion starts at $t=t_{\rm snow}$.

Using these numerical results,
we derived an analytical formula for $f_{\rm water}$ by icy pebble accretion.
In the formula, $f_{\rm water}$ is explicitly given as a function of
the ratio $t_{\rm snow}/t_{\rm pff}$ and the disk parameters.
The parameter $t_{\rm snow}/t_{\rm pff}$ is also determined by the disk parameters.
As a result, $f_{\rm water}$ is predicted by the disk parameters, especially
the initial disk mass accretion rate $\dot{M}_{\rm g,0}$ and initial disk size $\rd0$.
 It is insensitive to
the pebble accretion parameters such as the planet mass and Stokes number of 
drifting pebbles.

We found that
the expected water fraction of an Earth analogue near 1 au 
has $f_{\rm water} \sim 10^{-4}-10^{-2}$, which may correspond to
the value of the current Earth, 
in the disks with initial disk size of $r_{\rm d,0} \sim$ 30-50 au and
the initial disk mass accretion rate $\dot{M}_{\rm g,0} \sim (10^{-8}-10^{-7}) M_\odot{\rm/r}$.
For $\dot{M}_{\rm g,0} \ga 2 \times 10^{-8} M_\odot{\rm/r}$,
both the Earth and a Mars analogues have $f_{\rm water} \sim 10^{-4}-10^{-2}$,
while $f_{\rm water}$ is generally larger for Mars than for Earth.
Because these disks may be median or slightly compact/massive disks 
among  classical T Tauri stars,
our results suggest that rocky planets in habitable zones 
in exoplanatery systems around solar-type stars often 
have the water fraction similar to the Earth, if the pebble accretion is responsible for the water delivery.

\begin{acknowledgements}
We thank Michiel Lambrechts for detailed and helpful comments.
This work was supported by JSPS KAKENHI 15H02065
and 16K17661 and by MEXT KAKENHI 18H05438.
\end{acknowledgements}

\bibliographystyle{aa}


\begin{thebibliography}{57}
\expandafter\ifx\csname natexlab\endcsname\relax\def\natexlab#1{#1}\fi

\bibitem[{{Armitage} {et~al.}(2013){Armitage}, {Simon}, \&
  {Martin}}]{Armitage2013}
{Armitage}, P.~J., {Simon}, J.~B., \& {Martin}, R.~G. 2013, \apjl, 778, L14

\bibitem[{{Ataiee} {et~al.}(2018){Ataiee}, {Baruteau}, {Alibert}, \&
  {Benz}}]{Ataiee18}
{Ataiee}, S., {Baruteau}, C., {Alibert}, Y., \& {Benz}, W. 2018, \aap, 615,
  A110

\bibitem[{{Bai} {et~al.}(2016){Bai}, {Ye}, {Goodman}, \& {Yuan}}]{Bai16}
{Bai}, X.-N., {Ye}, J., {Goodman}, J., \& {Yuan}, F. 2016, \apj, 818, 152

\bibitem[{{Bercovici} \& {Karato}(2003)}]{BandK2003}
{Bercovici}, D. \& {Karato}, S.-i. 2003, \nat, 425, 39

\bibitem[{{Birnstiel} {et~al.}(2012){Birnstiel}, {Klahr}, \&
  {Ercolano}}]{Birnstiel2012}
{Birnstiel}, T., {Klahr}, H., \& {Ercolano}, B. 2012, \aap, 539, A148

\bibitem[{{Bitsch} {et~al.}(2018){Bitsch}, {Morbidelli}, {Johansen}, {Lega},
  {Lambrechts}, \& {Crida}}]{Bitsch2018}
{Bitsch}, B., {Morbidelli}, A., {Johansen}, A., {et~al.} 2018, \aap, 612, A30

\bibitem[{{Brauer} {et~al.}(2008){Brauer}, {Dullemond}, \&
  {Henning}}]{Brauer2008}
{Brauer}, F., {Dullemond}, C.~P., \& {Henning}, T. 2008, \aap, 480, 859

\bibitem[{{Brown}(1949)}]{Brown49}
{Brown}, H. 1949, in The Atmospheres of the Earth and Planets, ed. G.~P.
  {Kuiper}, 258

\bibitem[{{Clifford} {et~al.}(2010){Clifford}, {Lasue}, {Heggy}, {Boisson},
  {McGovern}, \& {Max}}]{Clifford2010}
{Clifford}, S.~M., {Lasue}, J., {Heggy}, E., {et~al.} 2010, Journal of
  Geophysical Research (Planets), 115, E07001

\bibitem[{{di Achille} \& {Hynek}(2010)}]{AchilleHynek2010}
{di Achille}, G. \& {Hynek}, B.~M. 2010, in Lunar and Planetary Science
  Conference, Vol.~41, Lunar and Planetary Science Conference, 2366

\bibitem[{{Donahue} {et~al.}(1982){Donahue}, {Hoffman}, {Hodges}, \&
  {Watson}}]{Donahue1982}
{Donahue}, T.~M., {Hoffman}, J.~H., {Hodges}, R.~R., \& {Watson}, A.~J. 1982,
  Science, 216, 630

\bibitem[{Fei {et~al.}(2017)Fei, Yamazaki, Sakurai, Miyajima, Ohfuji, Katsura,
  \& Yamamoto}]{Fei2017}
Fei, H., Yamazaki, D., Sakurai, M., {et~al.} 2017, Science Advances, 3,
  e1603024

\bibitem[{{Garaud} \& {Lin}(2007)}]{GaraudLin2007}
{Garaud}, P. \& {Lin}, D.~N.~C. 2007, \apj, 654, 606

\bibitem[{{Genda} \& {Abe}(2005)}]{GendaAbe2005}
{Genda}, H. \& {Abe}, Y. 2005, \nat, 433, 842

\bibitem[{{Greenwood} {et~al.}(2018){Greenwood}, {Karato}, {Vander Kaaden},
  {Pahlevan}, \& {Usui}}]{Greenwood2018}
{Greenwood}, J.~P., {Karato}, S.-i., {Vander Kaaden}, K.~E., {Pahlevan}, K., \&
  {Usui}, T. 2018, \ssr, 214, 92

\bibitem[{{Guillot} {et~al.}(2014){Guillot}, {Ida}, \& {Ormel}}]{Guillot2014}
{Guillot}, T., {Ida}, S., \& {Ormel}, C.~W. 2014, \aap, 572, A72

\bibitem[{{Gundlach} \& {Blum}(2015)}]{GundlachBlum15}
{Gundlach}, B. \& {Blum}, J. 2015, \apj, 798, 34

\bibitem[{{Haisch} {et~al.}(2001){Haisch}, {Lada}, \& {Lada}}]{Haisch2001}
{Haisch}, Jr., K.~E., {Lada}, E.~A., \& {Lada}, C.~J. 2001, \apjl, 553, L153

\bibitem[{{Hartmann} {et~al.}(1998){Hartmann}, {Calvet}, {Gullbring}, \&
  {D'Alessio}}]{Hartmann1998}
{Hartmann}, L., {Calvet}, N., {Gullbring}, E., \& {D'Alessio}, P. 1998, \apj,
  495, 385

\bibitem[{{Hartmann} {et~al.}(2016){Hartmann}, {Herczeg}, \&
  {Calvet}}]{Hartmann2016}
{Hartmann}, L., {Herczeg}, G., \& {Calvet}, N. 2016, \araa, 54, 135

\bibitem[{{Hasegawa} {et~al.}(2017){Hasegawa}, {Okuzumi}, {Flock}, \&
  {Turner}}]{Hasegawa2017}
{Hasegawa}, Y., {Okuzumi}, S., {Flock}, M., \& {Turner}, N.~J. 2017, \apj, 845,
  31

\bibitem[{{Hirose} \& {Turner}(2011)}]{HiroseTurner11}
{Hirose}, S. \& {Turner}, N.~J. 2011, \apjl, 732, L30

\bibitem[{{Hirschmann}(2006)}]{Hirschmann2006}
{Hirschmann}, M.~M. 2006, Annual Review of Earth and Planetary Sciences, 34,
  629

\bibitem[{{Ida} {et~al.}(2016){Ida}, {Guillot}, \& {Morbidelli}}]{Ida-et2016}
{Ida}, S., {Guillot}, T., \& {Morbidelli}, A. 2016, \aap, 591, A72

\bibitem[{{Johansen} {et~al.}(2018){Johansen}, {Ida}, \&
  {Brasser}}]{Johansen2018}
{Johansen}, A., {Ida}, S., \& {Brasser}, R. 2018, arXiv e-prints
  [\eprint[arXiv]{1811.00523}]

\bibitem[{{Klahr} {et~al.}(2018){Klahr}, {Pfeil}, \& {Schreiber}}]{Klahr18}
{Klahr}, H., {Pfeil}, T., \& {Schreiber}, A. 2018, {Instabilities and Flow
  Structures in Protoplanetary Disks: Setting the Stage for Planetesimal
  Formation}, 138

\bibitem[{{Kruijer} {et~al.}(2017){Kruijer}, {Burkhardt}, {Budde}, \&
  {Kleine}}]{Kruijer2017}
{Kruijer}, T.~S., {Burkhardt}, C., {Budde}, G., \& {Kleine}, T. 2017,
  Proceedings of the National Academy of Science, 114, 6712

\bibitem[{{Kurokawa} {et~al.}(2014){Kurokawa}, {Sato}, {Ushioda}, {Matsuyama},
  {Moriwaki}, {Dohm}, \& {Usui}}]{Kurokawa2014}
{Kurokawa}, H., {Sato}, M., {Ushioda}, M., {et~al.} 2014, Earth and Planetary
  Science Letters, 394, 179

\bibitem[{{Lambrechts} \& {Johansen}(2012)}]{OrmelKlahr2012}
{Lambrechts}, M. \& {Johansen}, A. 2012, \aap, 544, A32

\bibitem[{{Lambrechts} \& {Johansen}(2014)}]{LJ2014}
{Lambrechts}, M. \& {Johansen}, A. 2014, \aap, 572, A107

\bibitem[{{Lambrechts} {et~al.}(2014){Lambrechts}, {Johansen}, \&
  {Morbidelli}}]{Lambrechts2014}
{Lambrechts}, M., {Johansen}, A., \& {Morbidelli}, A. 2014, \aap, 572, A35

\bibitem[{{Lunine} {et~al.}(2003){Lunine}, {Chambers}, {Morbidelli}, \&
  {Leshin}}]{Lunine2003}
{Lunine}, J.~I., {Chambers}, J., {Morbidelli}, A., \& {Leshin}, L.~A. 2003,
  \icarus, 165, 1

\bibitem[{{Lynden-Bell} \& {Pringle}(1974)}]{LyndenPringle1974}
{Lynden-Bell}, D. \& {Pringle}, J.~E. 1974, \mnras, 168, 603

\bibitem[{{Lyra} \& {Umurhan}(2018)}]{LyraUmurhan18}
{Lyra}, W. \& {Umurhan}, O. 2018, arXiv e-prints [\eprint[arXiv]{1808.08681}]

\bibitem[{{Machida} \& {Abe}(2010)}]{MachidaAbe2010}
{Machida}, R. \& {Abe}, Y. 2010, \apj, 716, 1252

\bibitem[{{Marty}(2012)}]{Marty2012}
{Marty}, B. 2012, Earth and Planetary Science Letters, 313, 56

\bibitem[{{Matsumura} {et~al.}(2016){Matsumura}, {Brasser}, \&
  {Ida}}]{Matsumura2016}
{Matsumura}, S., {Brasser}, R., \& {Ida}, S. 2016, \apj, 818, 15

\bibitem[{{Matsumura} {et~al.}(2017){Matsumura}, {Brasser}, \&
  {Ida}}]{Matsumura2017}
{Matsumura}, S., {Brasser}, R., \& {Ida}, S. 2017, \aap, 607, A67

\bibitem[{{Min} {et~al.}(2011){Min}, {Dullemond}, {Kama}, \&
  {Dominik}}]{Min2011}
{Min}, M., {Dullemond}, C.~P., {Kama}, M., \& {Dominik}, C. 2011, \icarus, 212,
  416

\bibitem[{{Morbidelli} {et~al.}(2016){Morbidelli}, {Bitsch}, {Crida},
  {Gounelle}, {Guillot}, {Jacobson}, {Johansen}, {Lambrechts}, \&
  {Lega}}]{Morbidelli2016}
{Morbidelli}, A., {Bitsch}, B., {Crida}, A., {et~al.} 2016, \icarus, 267, 368

\bibitem[{{Morbidelli} {et~al.}(2000){Morbidelli}, {Chambers}, {Lunine},
  {Petit}, {Robert}, {Valsecchi}, \& {Cyr}}]{Morbidelli2000}
{Morbidelli}, A., {Chambers}, J., {Lunine}, J.~I., {et~al.} 2000, Meteoritics
  and Planetary Science, 35, 1309

\bibitem[{{Musiolik} {et~al.}(2016){Musiolik}, {Teiser}, {Jankowski}, \&
  {Wurm}}]{Musiolik16}
{Musiolik}, G., {Teiser}, J., {Jankowski}, T., \& {Wurm}, G. 2016, \apj, 818,
  16

\bibitem[{{Nomura} {et~al.}(2014){Nomura}, {Hirose}, {Uesugi}, {Ohishi},
  {Tsuchiyama}, {Miyake}, \& {Ueno}}]{Nomura2014}
{Nomura}, R., {Hirose}, K., {Uesugi}, K., {et~al.} 2014, Science, 343, 522

\bibitem[{{O'Brien} {et~al.}(2014){O'Brien}, {Walsh}, {Morbidelli}, {Raymond},
  \& {Mandell}}]{OBrien2014}
{O'Brien}, D.~P., {Walsh}, K.~J., {Morbidelli}, A., {Raymond}, S.~N., \&
  {Mandell}, A.~M. 2014, \icarus, 239, 74

\bibitem[{{Oka} {et~al.}(2011){Oka}, {Nakamoto}, \& {Ida}}]{Oka2011}
{Oka}, A., {Nakamoto}, T., \& {Ida}, S. 2011, \apj, 738, 141

\bibitem[{{Okuzumi} {et~al.}(2012){Okuzumi}, {Tanaka}, {Kobayashi}, \&
  {Wada}}]{Okuzumi2012}
{Okuzumi}, S., {Tanaka}, H., {Kobayashi}, H., \& {Wada}, K. 2012, \apj, 752,
  106

\bibitem[{{Ormel}(2014)}]{Ormel2014}
{Ormel}, C.~W. 2014, \apjl, 789, L18

\bibitem[{{Ormel} \& {Klahr}(2010)}]{OrmelKlahr2010}
{Ormel}, C.~W. \& {Klahr}, H.~H. 2010, \aap, 520, A43

\bibitem[{{Ormel} \& {Kobayashi}(2012)}]{OrmelKobayashi2012}
{Ormel}, C.~W. \& {Kobayashi}, H. 2012, \apj, 747, 115

\bibitem[{{Raymond} {et~al.}(2004){Raymond}, {Quinn}, \&
  {Lunine}}]{Raymond2004}
{Raymond}, S.~N., {Quinn}, T., \& {Lunine}, J.~I. 2004, \icarus, 168, 1

\bibitem[{{Sato} {et~al.}(2016){Sato}, {Okuzumi}, \& {Ida}}]{Sato2016}
{Sato}, T., {Okuzumi}, S., \& {Ida}, S. 2016, \aap, 589, A15

\bibitem[{{Shakura} \& {Sunyaev}(1973)}]{Shakura1973}
{Shakura}, N.~I. \& {Sunyaev}, R.~A. 1973, \aap, 24, 337

\bibitem[{{Suzuki} {et~al.}(2016){Suzuki}, {Ogihara}, {Morbidelli}, {Crida}, \&
  {Guillot}}]{Suzuki16}
{Suzuki}, T.~K., {Ogihara}, M., {Morbidelli}, A., {Crida}, A., \& {Guillot}, T.
  2016, \aap, 596, A74

\bibitem[{{Takeuchi} \& {Lin}(2005)}]{TakeuchiLin2005}
{Takeuchi}, T. \& {Lin}, D.~N.~C. 2005, \apj, 623, 482

\bibitem[{{Wada} {et~al.}(2009){Wada}, {Tanaka}, {Suyama}, {Kimura}, \&
  {Yamamoto}}]{Wada09}
{Wada}, K., {Tanaka}, H., {Suyama}, T., {Kimura}, H., \& {Yamamoto}, T. 2009,
  \apj, 702, 1490

\bibitem[{{Williams} \& {Cieza}(2011)}]{WilliamsCieza2011}
{Williams}, J.~P. \& {Cieza}, L.~A. 2011, \araa, 49, 67

\bibitem[{{Youdin} \& {Lithwick}(2007)}]{Youdin2007}
{Youdin}, A.~N. \& {Lithwick}, Y. 2007, \icarus, 192, 588

\end{thebibliography}

\end{document}